\documentclass[fleqn,usenatbib]{mnras}
\usepackage{anyfontsize}
\usepackage{newtxtext,newtxmath}
\usepackage[T1]{fontenc}

\DeclareRobustCommand{\VAN}[3]{#2}
\let\VANthebibliography\thebibliography
\def\thebibliography{\DeclareRobustCommand{\VAN}[3]{##3}\VANthebibliography}

\usepackage{graphicx}	% Including figure files
\usepackage{amsmath}	% Advanced maths commands
\graphicspath{{plots/}}
\usepackage{subcaption}
\usepackage{multicol}
\usepackage{multirow}
\usepackage{xspace}
\usepackage{comment}
\usepackage{xcolor}
\usepackage{natbib}
\usepackage{threeparttable}
\usepackage{lscape}
\usepackage{booktabs}
\usepackage{xcolor,colortbl}
\definecolor{lavender}{rgb}{0.9, 0.9, 0.98}

\newcommand{\pinocchio}{\texttt{PINOCCHIO}\xspace}
\newcommand{\gaea}{\textsc{gaea}\xspace}
\title[Host galaxies of Pop III.1 seeded SMBHs]{The formation of supermassive black holes from Population III.1 seeds. III. Galaxy evolution and black hole growth from semi-analytic modelling}

\author[Cammelli et al.]{Vieri Cammelli,$^{1,2,3,4}$\thanks{E-mail: vieri.cammelli@phd.units.it} 
Pierluigi Monaco,$^{1,2,4,6}$
Jonathan C. Tan,$^{3,5}$
Jasbir Singh,$^{7}$
$\;$Fabio Fontanot,$^{2,4}$
\newauthor
Gabriella De Lucia,$^{2,4}$
Michaela Hirschmann,$^{2,8}$
and Lizhi Xie,$^{2,9}$
\\
$^{1}$Astronomy Unit, Department of Physics, University of Trieste, via G.B. Tiepolo 11, I-34143, Trieste, Italy\\
$^{2}$INAF - Astronomical Observatory of Trieste, via G.B. Tiepolo 11, I-34143, Trieste, Italy\\
$^{3}$Department of Space, Earth \& Environment, Chalmers University of Technology, SE-412 96 Gothenburg, Sweden\\
$^{4}$IFPU – Institute for Fundamental Physics of the Universe, Via Beirut 2, I-34151 Trieste, Italy\\
$^{5}$Dept. of Astronomy, University of Virginia, Charlottesville, VA 22904, USA\\
$^{6}$INFN, Sezione di Trieste, Via Valerio 2, I-34127 Trieste, Italy\\
$^{7}$INAF – Astronomical Observatory of Brera, via Brera 28, I-20121 Milan, Italy\\
$^{8}$Institute of Physics, Lab for Galaxy Evolution, EPFL, Observatoire de Sauverny, Chemin Pegasi 51, 1290 Versoix, Switzerland\\
$^{9}$Tianjin Normal University, Binshuixidao 393, Xiqing, 300387, Tianjin, People's Republic of China
}

\date{Accepted 2024 November 26. Received 2024 October 28; in original form 2024 July 13}

\pubyear{2024}

\begin{document}
\label{firstpage}
\pagerange{\pageref{firstpage}--\pageref{lastpage}}
\maketitle

\begin{abstract}
We present an implementation of Pop III.1 seeding of supermassive black holes (SMBHs) in a theoretical model of galaxy formation and evolution to assess the growth of the SMBH population and the properties of the host galaxies. The model of Pop III.1 seeding involves SMBH formation at redshifts $z\gtrsim 20$ in dark matter minihalos that are isolated from external radiative feedback, parameterized by isolation distance $d_{\rm iso}$.
Within a standard $\Lambda$CDM cosmology, we generate dark matter halos using the code \pinocchio and seed them according to the Pop III.1 scenario, exploring values of $d_{\rm iso}$ from 50 to 100~kpc (proper distance). We consider two alternative cases of SMBH seeding: a Halo Mass Threshold (HMT) model in which all halos $>7\times10^{10}\:M_\odot$ are seeded with $\sim 10^5\:M_\odot$ black holes; an All Light Seed (ALS) model in which all halos are seeded with low, stellar-mass black holes. 
We follow the redshift evolution of the halos, populating them with galaxies using the GAlaxy Evolution and Assembly theoretical model of galaxy formation, including accretion on SMBHs and related feedback processes.  
Here we present predictions for the properties of galaxy populations, focusing on stellar masses, star formation rates, and black hole masses. The local, $z\sim0$ metrics of occupation fraction as a function of the galaxy stellar mass, galaxy stellar mass function (GSMF), and black hole mass function (BHMF) all suggest a constraint of $d_{\rm iso}<75\:$kpc. We discuss the implications of this result for the Pop III.1 seeding mechanism.
%either all halos are seeded with lower-mass black holes or the seeds are assigned to all halos above a fixed mass threshold are seeded with BH masses around 100,000 solar masses.
%We consider different scenarios, among which the model arguing that the first (Pop III.1) isolated stars in the universe, forming at $z>20$ in dark matter minihalos, are affected by dark matter annihilation heating that allows them to grow to ~100,000 solar masses and become the progenitors of all SMBHs.
%Therefore, we can in principle disentangle the degeneracy among several seeding mechanisms by comparing to the observed SMBHs and host galaxies and investigating their impact especially in terms of observational quantities.
%This approach connects the merger histories of dark matter halos directly to the associated galaxy properties. 
%A future, companion paper presents the results for observable luminosities of the sources.
%In particular, we investigate how the presence of a SMBH appearing as an Active Galactic Nucleus (AGN) impact physical galaxy properties such as star formation rate, stellar mass, BH mass. 
% We examine how these model predictions may be tested by both current and next generation facilities like HST, JWST, E-ELT, Euclid \& ATHENA.
\end{abstract}


%%%%%%%%%%%%%%%%%%%%%%%%%%%%%%%%%%%%%%%%%%%%%%%%%%

%%%%%%%%%%%%%%%%% BODY OF PAPER %%%%%%%%%%%%%%%%%%

%\pigi{NB: pedices in math mode must be in rm (see d\_iso below).}

% Vieri, please check the consistency of all acronyms

\section{Introduction}
\label{sec:intro}
The origin of supermassive black holes (SMBHs) that reside in the nuclei of massive galaxies is one of the most pressing topics of current astrophysics research.
%spanning over four decades of proposed mechanisms (e.g., \citet{Rees78}). 
A central challenge lies in explaining the large masses of the black holes observed in bright quasars (QSOs) at early cosmic epochs. For instance, \citet{Wang21} report a $\sim10^{9}\ M_{\odot}$ SMBH at $z = 7.642$ which, even assuming Eddington limited accretion throughout its entire lifetime, would require a seed mass of $\sim10^{4} M_{\odot}$ at $z\sim30$. These observations impose stringent constraints on theories of SMBH formation and growth. 

How SMBHs are seeded has been the subject of numerous studies \citep[see reviews of, e.g.,][]{Rees78,Volonteri10,Inayoshi20}.
%(see reviews of \citet{Volonteri10} and \citet{Inayoshi20}). 
Among proposed mechanisms, the \textit{direct collapse} (DCBH) scenario advances the idea that a primordial gas cloud hosted by a relatively massive, atomically-cooled, UV-irradiated halo of $\sim10^{8} M_{\odot}$ collapses into a single supermassive star of $\sim10^{4-6} M_{\odot}$, subsequently forming a \textit{massive seed} by $z\sim10$ (e.g., \citet{Bromm03, Begelman06, Montero12}). While this mechanism can explain the number density of high redshift quasars, its strict conditions hinder the formation of sufficient numbers of SMBHs to account for the observed SMBH population at $z=0$ \citep{Chon16, Wise19}. Other works indicate \textit{light} seeds, of the order of 100 $M_{\odot}$, to be byproducts of ``standard'' Pop III stars forming in $\sim10^6\:M_\odot$ dark matter ``minihalos'' at $z\sim20$ (e.g., \citet{Madau01,2004ApJ...603..383T,McKee08}). Moreover, very dense star clusters at high $z$ may dynamically evolve leading to the formation of \textit{intermediate}-mass black hole (IMBH) seeds with $10^{2-4} M_{\odot}$ (e.g., \citet{Portegies04, Devecchi09}). 
%While a combination of these formation scenarios may explain some of the observed features, this has not been explored yet 

%given the complexity of implementing such different phenomena in a comprehensive way. 
One challenge facing models that start with light or intermediate-mass seeds is the apparent dearth of observed intermediate-mass black holes (IMBHs) in the local universe
%Yet, none of these mechanisms alone can consistently reproduce the whole population of SMBHs and explain the dearth of observations of intermediate mass BHs
\citep{Banik19,Greene20,Volonteri21}. In addition, it is difficult to model the IMBH seeding in a cosmological context, unless requiring either sub-grid or probabilistic recipes calibrated on high-resolution zoom simulations as recently studied by \cite{Bhowmick24a, Bhowmick24b}.

An alternative scenario of SMBH formation from Pop III.1 protostars has been studied in a cosmological context by \citet{Banik19} (hereafter Paper I) and \citet{Singh23} (hereafter Paper II). Pop III.1 sources are defined as special Pop III stars forming at the center of dark matter (DM) minihalos in the early universe ($z\gtrsim20$), which are isolated from any source of stellar or SMBH feedback \citep{McKee08}. This criterion is parameterized by an isolation distance $d_{\rm iso}$, expressed in physical units, of the order of $\lesssim$100 kpc. The physical mechanism that allows growth of Pop III.1 protostars to high mass, i.e., $\gtrsim 10^4\:M_\odot$, is the influence of dark matter annihilation (DMA) on the protostellar structure \citep{Spolyar08,Natarajan09,Freese10,Rindler-Daller15}. The fiducial assumption is that dark matter is composed of a weakly interacting massive particles (WIMPs), which undergoes self-annihilation with a weak interaction cross-section. If sufficient WIMPs are captured by the Pop III.1 protostar, then its structure is altered. In particular, the protostar can remain relatively large as it accretes to high masses, thus reducing its ionizing feedback on its own accretion flow. This may enable the efficient accretion of a large fraction of the entire baryonic content of the parent minihalo, i.e., $\sim10^{5} M_{\odot}$, to the Pop III.1 protostar, which subsequently, within a few Myr, collapses to a SMBH. Other minihalos are Pop III.2 sources, i.e., still metal free, but, having been irradiated by UV radiation, have higher free electron abundances leading to greater abundances of $\rm H_2$ and $\rm HD$, higher cooling rates, and fragmentation to lower mass, $\sim 10\:M_\odot$, stars \citep{2006MNRAS.366..247J,2006MNRAS.373..128G}.

The Pop III.1 model thus predicts a characteristic mass of $\sim10^{5} M_{\odot}$ for SMBH seeds and provides an explanation for the apparent dearth of IMBHs. In addition, the Pop III.1 model predicts that SMBHs form very early in the universe, which provides a theoretical explanation for the population of high-z SMBHs without the need for sustained Eddington or super-Eddington levels of accretion. Another prediction is that the initial spatial distribution of SMBHs is relatively uniform, i.e., with low levels of clustering, with the seeds separated from each other by distances of order $d_{\rm iso}$. The co-moving number density of SMBHs, $n_{\rm SMBH}$, is sensitive to $d_{\rm iso}$, with a value of about 100~kpc (proper distance) able to explain the observed local number density of SMBHs of $\sim 5\times 10^{-3}\:{\rm Mpc}^{-3}$. The value of $d_{\rm iso}$ also sets a limit on the formation epoch of SMBHs. For $d_{\rm iso}=50-100\:$kpc, most seeds are formed at $z\sim 30$, i.e., with co-moving separations of $\sim few\:$Mpc, and the process is largely complete by $z\sim 25$, after which $n_{\rm SMBH}$ is nearly constant. This time scale has a weak dependence on $d_{\rm iso}$, with later seeding happening for smaller $d_{\rm iso}$. As a result of the initially separated distributions, in the Pop III.1 model mergers only begin to occur at relatively late times, i.e., $z\lesssim 1$, which causes $n_{\rm SMBH}$ to decrease by modest amounts $\sim 10-20\%$ \citep{Singh23}. Consequently, the SMBH number density remains fairly constant down to the local Universe, with only a small fraction of seeds lost in halo mergers by $z=0$. At redshifts $> 5$, this model makes strong predictions for the number of SMBHs we expect in the early Universe (e.g., their relative contribution to the luminosity functions), underscoring the importance of current and future deep field observations of high redshift AGNs. Such observations are crucial for potentially distinguishing among different seeding mechanisms and/or accretion schemes. Furthermore, by redshift 0 the occupation fraction of seeded halos saturates to unity for the most massive halos as a function of the isolation distance $d_{\rm iso}$.

Numerical simulations of galaxy formation and evolution in cosmological volumes typically have quite limited spatial resolution and so struggle to capture the processes leading to SMBH seeding. The lack of knowledge about the physical processes and the difficulty to treat them from first principles further limit such numerical approaches. In addition, the real challenge resides in simultaneously simulating a broad range of scales, from mini-halos at $z\sim20-30$ up to the structures we observe in the local Universe. Thus, implemented seeding schemes are generally based on simple threshold models.
%From the numerical point of view, the mechanisms that drive BH formation and its first phases of accretion take places at very small scales, compared to the scales relevant for the formation of the host dark matter halo, hence simulation codes must rely on sub-resolution assumptions for the emergence of the SMBH seed and its accretion, which are generally based on the host halo/galaxy properties. For instance, when running large cosmological volumes, 
In particular, many works seed a SMBH once the parent halo reaches a certain threshold in dark matter mass, i.e., a halo mass threshold (HMT) seeding scheme \citep{Sijacki07,DiMatteo08}. The same idea has been used by \citet{Vogelsberger14} within the Illustris Project, as well as in the Evolution and Assembly of GaLaxies and their Environments (EAGLE) simulations of \citet{Barber16}. 

%Alternative efforts incorporate additional galaxy properties into SMBH seeding process. 

Extending to alternative threshold models, in the Horizon-active galactic nucleus (AGN) simulation (with DM mass resolution of $8\times10^8 M_{\odot}$), \citet{Volonteri16} implemented lower limits for gas and stellar densities, as well as stellar velocity dispersion, to determine if a galaxy  hosts a black hole, using a seed mass of $10^{5} M_{\odot}$. Their formation was restricted to redshifts $z>1.5$ and all forming black holes had to be separated by at least 50 comoving kiloparsecs to prevent multiple black holes from forming in the same galaxy. Similarly, the OBELISK simulation (\citet{Trebitsch21}, based on a sub-volume of the HORIZON-AGN simulation) used a slightly lower seed mass and applied gas and stellar density thresholds, including gas Jeans instability, requiring an isolation of 50 kpc from other SMBHs to prevent multiple black hole formation. In another approach, the ROMULUS simulation (\citet{Tremmel17}, particle DM mass $\sim3\times10^5 M_{\odot}$) set criteria based on metallicity limits, gas density thresholds, and a restricted temperature range for SMBH formation, using a seed mass of $10^{6} M_{\odot}$. Additionally, in the Illustris Project framework, \citet{Bhowmick22} explored various gas-based SMBH seeding prescriptions and a range of seed masses from around $10^{4}$ to $10^{6} M_{\odot}$, while keeping a relatively high DM resolution about $\sim6\times10^6 M_{\odot}$. Moreover, in high-resolution zoom simulations \cite{Bhowmick24a, Bhowmick24b, Bhowmick24c} explicitly resolved pristine and dense gas clouds forming $\sim10^{3} M_{\odot}$ seeds and built a stochastic seeding scheme that can directly set the initial seeding conditions in lower resolution runs. Using the hydrodynamical cosmological code RAMSES , \cite{Habouzit16} investigated the conditions under which halos can host DCBHs over a large range of resolutions and box sizes as a function of the illuminating Lyman-Werner (LW) background and supernova (SN) feedback. Under optimistic assumptions, their SMBH number density ranges from $7\times 10^{-7}$ up to $10^{-4}\:{\rm cMpc}^{-3}$, still a factor of at least $\sim10$ lower than the local estimate. Despite the improvements in the treatment of sub-grid physics and the implementation of zoom-in approaches, large, cosmological simulations do not include yet a full physical model for SMBH formation \citep[e.g., see][]{DiMatteo23}. 

A complementary approach with respect to hydrodynamical simulations is provided by semi-analytic models (SAMs). SAMs are tools used to simulate the evolution of galaxy populations within DMHs by modelling the physical processes that drive the evolution of the baryonic components of dark matter halos by invoking theoretically and/or observationally motivated prescriptions. We stress here that such prescriptions (\textit{sub-grid} physics) are similarly implemented in hydrodynamical simulations. These processes encompass baryonic gas cooling and heating, star formation, gas accretion onto SMBHs, and their related feedback mechanisms. The flexibility of these models permits us to obtain predictions of galaxy properties across cosmological volumes and allows efficient exploration of the associated parameter space in order to study the impact of different physical assumptions (for a review see \citet{Somerville15, DeLucia19b}). 
%Among them, the treatment of BH formation and dynamics has been studied in many SAM works (\citet{Sesana07, Volonteri09, Barausse12, , Valiante18, Dayal19, Sassano21}). 
On the other hand, the price to pay is the lack of a complete and consistent treatment of the gas hydrodynamics. One limitation of the application of SAMs to the study of the early BH seeding lies in the lack of resolution in dark matter halo trees generated with either N-body simulation or analytic recipes such as \cite{Press_Schechter1974} and \cite{LaceyCole93}. In fact, dealing with cosmological volumes significantly impacts the modelling of BH seed formation, which depends on the local gas conditions within halos and the DM mass resolution.

Among the BH seeding mechanisms implemented in SAMs, \cite{RicarteNatarajan18} used DCBHs whose mass is calculated proportionally to the DM halo mass. Given their redshift-dependent mass resolution scheme, this results in placing $\gtrsim 10^{4} M_{\odot}$ BH seeds at $z\sim 15-20$ in parent halos of minimum mass about $5\times10^{6} M_{\odot}$. In the Cosmic Archaeology Tool (CAT) presented in \cite{Trinca22}, both light seeds from Pop III stars and DCBHs are considered to study their contribution the BH mass function. Similarly, BH seed masses ranging from $10^{2}$ up to $10^{5} M_{\odot}$ were assigned  in the DELPHI SAM \citep{Dayal19} according to the probability of a halo to host a DCBH or a stellar BH remnant. \cite{Sassano21} seeded BHs at the center of galaxies based on the locally derived properties of the halo environment. In particular, several thresholds for the illuminating LW flux, the metallicity and the gas-to-dust ratio were used to determine whether the final BH would be a light, medium or heavy seed. In the L-GALAXIES model, \citep{Spinoso22} studied DCBHs according to spatial variations of the star formation in terms of chemical and radiative feedback. Since the first mini-halos ($T_{vir}\sim 10^3 K$) lie below their resolution limit, Pop III remnant seeds were placed following the sub-grid approach of \cite{Sassano21}.
%, while at intermediate ranges of redshift the fraction of seeded halos seems to be more uncertain. 

In this paper, we model SMBH formation in a cosmological box within the standard $\Lambda$CDM cosmogony by exploiting halo merger trees generated with \pinocchio  (\citet{Monaco02, Munari17}) to simulate the large-scale distribution of dark matter halos (DMHs). The \pinocchio code follows the evolution of cosmic perturbations and structures based on Lagrangian Perturbation Theory (LPT), ellipsoidal collapse and excursion-set theory and generates catalogs of collapsed objects (i.e., DMHs). Subsequently, these halos are seeded according to their isolation state at the formation time (see Paper I and Paper II). By adapting the \pinocchio DMH merger tree to a suitable format, we follow the redshift evolution of these structures and we populate them with galaxies utilizing the state-of-the-art GAlaxy Evolution and Assembly (\gaea) semi-analytic model (SAM) of galaxy formation and evolution (\citet{Hirschmann16, Fontanot20, DeLucia24}). Different seeding mechanisms can be explored in this framework, but in this study we focus mainly on the Pop III.1 model. 
%This work is the first step towards an investigation of how the presence of a SMBH correlates with galaxy properties in this model.
We note here that the two key tools that we use in our approach have been successfully tested against observational and numerical measurements: (i) the \gaea model is able to reproduce a wide range of observational results at $z=0$ and up to $z\sim5$; (ii) the high resolution realizations of halo merger trees constructed with \pinocchio result from an approximate method which has been shown to agree well with simulations down to $z=0$, as shown in the above cited papers and references therein. Note that \pinocchio merger trees converge with a better accuracy with respect to analytic-based ones whose implementation in recent works are not well tested below  z$\sim$4. 

% focuses on defining and comparing various approaches for black hole (BH) growth within theoretical models of galaxy formation.

This paper is organized as follows. In \S\ref{sec:methods}, we describe in detail our new fully semi-analytic approach and the implementation of SMBH seeding and accretion. The main results of this study, together with the possible implications, are presented in \S\ref{sec:results}. Finally we present our summary and conclusions in \S\ref{sec:conclusions}. A companion follow-up paper (Cammelli et al., in prep.) will present predictions for the luminosity functions of the galaxies and AGN and their comparison with observational data.

\section{Methods}
\label{sec:methods}

In this work, we couple DMH merger trees extracted from cosmological boxes simulated using the \pinocchio algorithm with the \gaea SAM, providing predictions of the properties of galaxy populations associated with the DMH distribution at various redshifts. Given the full semi-analytic approach, we can explore wide ranges of physical and/or observational properties of galaxies with flexibility in the choice of the parametrizations adopted for the physical processes at play. In the following we give a brief description of the \pinocchio runs and the \gaea model. We then describe how we interface the two codes. We design a specific code which takes the \pinocchio halos as inputs, models the physical properties of main and sub halos and returns DMH merger trees structurally equivalent to numerically derived ones. We focus our implementation on different assumptions for SMBH seeding mechanisms within the treatment of accretion modes onto the central black hole  already implemented in \gaea \citep{Fontanot20}.

In this paper we adopt the following nomenclature. Gravitationally bound DM structures that are not hosted by a larger bound structure are called \textit{halos}, the galaxies lying at their centers are called \textit{central galaxies}. 
Halos may host smaller bound clumps of DM, we call them \textit{subhalos}, and the galaxies at their center \textit{satellite galaxies}. In a simulation, the disruption of a subhalo may happen before the actual merger should take place, due to limited resolution. Therefore, its associated satellite can exist for some time after the disappearance of its subhalo; in this phase it is named \textit{orphan galaxy}.

%We adopt that main DM bound structures, \textit{halos}, host \textit{central} galaxies. We refer to DM structures belonging to another main structure as \textit{subhalos}, while the galaxies residing at their center are known as \textit{satellites}. Already existing satellite galaxies for which the hosting subhalo cannot be resolved at later times in the simulation are referred to as \textit{orphan} galaxies.

\subsection{The dark matter skeleton: \pinocchio}
\label{subsec:pinocchio}

\pinocchio \citep[PIN-pointing Orbit Crossing-Collapsed HIerarchical Objects]{Monaco02,Munari17} is a semi-analytic code that follows the formation and merger history of dark matter halos in Lagrangian space, that is the space defined by the initial positions of mass elements. One can think of \pinocchio as an algorithm applied to the initial conditions of a simulation, where the Lagrangian space is discretized into a grid and each cell is represented by a massive particle. First, an algorithm based on ellipsoidal collapse computes, for each particle, the time at which the particle is deemed to reach a multi-stream region (orbit crossing). The particles are then grouped into massive halos, whose position is estimated using LPT. As such, this code can be seen as a halo finder in Lagrangian space, that provides relatively accurate halo catalogs without running a full numerical simulation. While memory requirements are still high, so that a massive run requires a supercomputer, the computational cost is thousands time lower than an equivalent N-body simulation.

For this paper we use the \pinocchio run presented in Paper II of this series, \cite{Singh23}, i.e., a cubic box of side $59.7$ Mpc ($40$ $h^{-1}$ Mpc with $h=0.67$) with standard Planck cosmology (\citet{2020A&A...641A...6P}), sampled with $4096^3$ particles, for a particle mass of $1.23\times 10^5$ M$_\odot$. The smallest resolved halos were set at 10 particles (an acceptable value for a semi-analytic algorithm), i.e., $1.23\times 10^6\:M_\odot$. This box was processed in Paper II to compute which mini-halos host SMBH seeds in the Pop III.1 scenario, with fiducial seed mass assumed to be $10^5 M_\odot$. Distributing such a large box on hundreds of nodes makes it difficult to reconstruct massive halos, whose Lagrangian size may be similar or even exceed the size of a computational domain. As explained in Paper II, for evolution from $z=10$ down to $z=0$ the box was re-run at a lower resolution (using $1024^3$ particles) on a single node, thus avoiding any issue in the domain decomposition, and the information about which halos are seeded was propagated in Lagrangian space by assigning the seeds to the lower-resolution particle that contains the seeded mini-halo. In the low-resolution run (used in this work), the dark matter particle mass is $\sim5 \times 10^{7}M_\odot$, and the smallest resolved halo is a factor of 10 more massive. \pinocchio is also able to produce the merger history of dark matter halos with continuous time sampling by providing the exact redshift for each individual merger event.
%However, this merger history cannot be directly fed to the \gaea semi-analytic model, that is based on numerical merger trees with discrete time sampling and where the halo substructure is resolved.
%For the purposes of this study we run a high resolution (dark matter particle mass of $\sim5 \times 10^{7}$ \(M_\odot\)) \pinocchio box of $40$ Mpc $h^{-1}$.
%\\
%Pigi's duty ;)

\subsection{GAEA semi-analytic model }
\label{subsec:gaea}

\gaea represents an evolution of the original model published by \cite{DeLucia_Blaizot07}. In this study, we utilize the version of the model published in \cite[][hereafter F20]{Fontanot20}. The model includes: (a) a comprehensive treatment of chemical enrichment, explicitly addressing differential enrichment linked to asymptotic giant branch (AGB) stars, Type II SNe, and Type Ia SNe (\citet{DeLucia14}); (b) an updated approach to stellar feedback tracing  gas ejection via stellar-driven outflows (the model is partially based on results from hydrodynamical simulations, \citet{Hirschmann16}) coupled with a gas re-incorporation timescale dependent on DMH mass (\citet{Henriques15}); (c) an improved model for disc sizes (\citet{Xie17}) tracking angular momentum evolution through mass and energy exchanges within the galaxy; (d) an update modelling for cold gas accretion onto SMBHs (F20). This latter ingredient is relevant for this study and will be described in more detail in the next subsection.

\gaea has been shown to reproduce a wide range of observations. \cite{Fontanot17b} demonstrated that the evolution of the galaxy stellar mass function (GSMF) and cosmic star formation rate (SFR) obtained from \gaea are in agreement with measurements available up to $z \sim 7$. \cite{Hirschmann16}  showed that the model reproduces well the observed gas fractions and mass-metallicity relations at $z<3$, but tends to overpredict the SF activity of low-mass galaxies at low redshift. The model we use in this study also nicely reproduces the fraction of quiescent galaxies as a function of stellar mass and hierarchy at low-$z$ \citep{DeLucia19}. Furthermore, the model galaxies exhibit size–mass and angular momentum–mass relations that are in relatively good agreement with observational assessments, both in the local Universe and at higher redshifts \citep{Zoldan19}. In future work we will extend our analysis to alternate recent versions of the model including distinct treatments for the partitioning of cold gas into atomic and molecular hydrogen \citep{Xie17, Xie20, DeLucia24}, as well as a model accounting for a variable stellar initial mass function \citep{Fontanot17a, Fontanot18, Fontanot24}. However, these two variations will not be considered in the present study.
%as we will extend our analysis to more recent version of the model in future work.

\subsubsection{SMBH accretion and feedback}
\label{subsec:accretion}
\begin{comment}    
Regarding the SMBH activity scheme assumed in \gaea, it was originally derived from the models outlined in previous studies such as \citet{Croton06} and \citet{DeLucia_Blaizot07}, which are based on \citet{Kauffmann00}. However, it is noteworthy that this implementation has known limitations in replicating the spatial density of luminous AGNs at high redshifts (\citet{Marulli08}). 
Regarding the SMBH activity scheme assumed in this work, we follow the implementation of F20 in \gaea. We recall that in this model the growth of SMBHs is mainly driven by galaxy mergers, where a portion of cold gas is promptly accreted onto the SMBH. This fraction is contingent upon the mass ratio ($m_{\rm rat}$) between the merging galaxies, the total available cold gas ($M_{\rm cold}$), the virial velocity of the hosting dark matter halo ($V_{\rm vir}$), and is modulated by the free parameter $f_{\rm k}$:
\begin{equation} \label{eq:qso_mode}
    \dot{M}_{\text{Q}} = f_{\rm k} \frac{m_{\rm rat}M_{\rm cold}}{1 + (V_{\rm vir}/280[km/s])^{-2}}.
\end{equation}
\end{comment}

The modelling adopted in \gaea for the accretion onto the SMBH is described in detail in F20. We note here that this phenomenon is treated following two main specific prescriptions. A first accretion channel from hot gas, known as \textit{radio-mode}, is modelled according to the implementation of \cite{Croton06}. In this mode, the accretion rate is proportional to the mass of the BH ($M_{\rm BH}$), to the virial velocity $V_{\rm vir}$ and to the fraction of the hot gas in the DMH ($f_{\rm hot}$), adjusted by a free parameter $k_{\rm radio}$:
\begin{equation} \label{eq:radio_mode}
    \dot{M}_{\text{R}} = k_{\rm radio} \frac{M_{\rm BH}}{10^8 M_{\odot}} \frac{f_{\rm hot}}{0.1} \Bigg(\frac{V_{\rm vir}}{200\:{\rm km/s}}\Bigg)^3.
\end{equation}
\begin{comment}
It is worth underlining that, in \citet{Hirschmann16}, the free parameters in equations (\ref{eq:qso_mode}) and (\ref{eq:radio_mode}) have been calibrated to reproduce the evolution of the GSMF up to $z\sim3$ and the local relation between $M_{\rm BH}$ and $M_{\rm bulge}$. 
\end{comment}

However, the more luminous AGNs arise from a second accretion mode, traditionally termed the \textit{QSO-mode}: in particular we take advantage of the modeling of cold gas accretion onto SMBHs presented in F20. In particular, we refer to the model implementation defined as F06-GAEA in F20, which is based on prescriptions first described by \cite{Fontanot06}. The occurrence of the AGN phenomenon has been accomplished by using a three phase approach. 1) The first phase requires that a fraction of the cold gas available in the galaxy dissipates a substantial amount of angular momentum and gathers in the central region, turning into a gas reservoir available for accretion onto the BH. 2) The amount of cold gas flowing from the reservoir towards the centre of the galaxy leads to accretion onto the BH. 3) Ultimately, outflows induced by the AGN lead to the expulsion of a portion of the galaxy's gas content.

% \subsubsection{Gas reservoir accretion}
This model assumes that disc instabilities and galaxy mergers lead to an efficient angular momentum loss and trigger QSO-mode accretion events. Typically this loss of angular momentum results from SFR episodes in the central regions, which inject turbulence and exert radiation drag. We assume that, following merger events, the BH reservoir accretion rate is proportional to the central SFR via the free parameter $f_{\rm lowJ}$ schematized as:
\begin{equation} \label{eq:merg}
    \dot{M}_{\rm rsv}^{\rm cs} = f_{\rm lowJ} \psi_{\rm cs},
\end{equation}
with $\psi_{\rm cs}$ estimated via the collisional starburst prescriptions from \cite{Somerville01} and it equals the amount of SFR in the central regions triggered by the merger itself. It is important to note that BHs are assumed to merge instantaneously (no delay is assigned) once the host galaxies have merged.

For disk instabilities, the net result in \gaea involves moving a fraction of stars from the stellar disk to the stellar bulge so as to restore stability \citep{DeLucia11}. Since there is no star formation associated by construction, we assume the reservoir growth rate to be proportional to the bulge grow rate $\dot{M}_{\rm bulge}$. Hence:
\begin{equation}\label{eq:disc}
    \dot{M}_{\rm rsv}^{\rm di} = f_{\rm lowJ} \mu \dot{M}_{\rm bulge},
\end{equation}
where the free parameters $f_{\rm lowJ}$ and $\mu$ ($6\times10^{-3}$ and 10, respectively, as in F20) regulate the fraction of gas accreted due to angular momentum loss. 

% \subsubsection{Viscous accretion}
Once the reservoir gathers gas around the central BH, accretion episodes can be triggered. Following the viscous accretion rate derived by \cite{Granato04}, we define the accretion onto the BH as:
\begin{equation} \label{eq:rate_bh}
    \dot{M}_{\rm BH} = f_{\rm BH} \frac{\sigma_{\rm B}^3}{G} \Bigg(\frac{M_{\rm rsv}}{M_{\rm BH}}\Bigg)^{3/2} \Bigg(1 + \frac{M_{\rm BH}}{M_{\rm rsv}}\Bigg)^{1/2},
\end{equation}
where $\sigma_{\rm B}$ is the velocity dispersion of the bulge component, assumed to scale linearly with $V_{\rm vir}$ as derived by \cite{Ferrarese02} for a sample of local galaxies. 

This prescription, once coupled with the amount of gas accumulated into the reservoir, can induce accretion rates beyond the Eddington limit. We limit the actual accretion rate to:  
\begin{equation} \label{eq:rate_max}
    \dot{M}_{\rm max} = 10 \frac{M_{\rm BH}}{t_{\rm edd}} = 10 \dot{M}_{\rm edd},
\end{equation}
where $\dot{M}_{\rm edd}$ is the accretion rate of a BH shining at the Eddington luminosity with a radiative efficiency of $10\%$, over an Eddington-Salpeter timescale $t_{\rm edd} \simeq 45\:$Myr. 
This upper limit is supported by both observational and theoretical findings \citep[see, e.g.,][]{Takeo19, Jiang19, Delvecchio20}. Some theoretical models indicate that such large accretion rates may occur via intermittent bursts, particularly at higher redshifts \citep[e.g.,][]{Inayoshi16}.

%\subsubsection{AGN feedback}
F20 examined the impact of AGN activity on the host galaxy, specifically focusing on its cold gas phase. AGNs are believed to influence the surrounding medium by actively heating it, and eventually leading to the expulsion of cold gas via galactic winds driven by the AGN. SNe explosions combined with the radiation pressure of the AGN are assumed to promote further accretion by compressing part of the ISM ($f_{\rm cen} \sim 10^{-3}$) in the central region. This material is eventually added to the BH reservoir \citep[see][]{Monaco05}. Each accretion episode described by Eq.~\eqref{eq:rate_bh} triggers an AGN-driven outflow, with a rate that is modelled assuming a scaling relation with the BH accretion rate:
\begin{equation}
    \dot{M}_{\rm qw} = \epsilon_{\rm qw} \dot{M}_{\rm BH},
\end{equation}
where $\epsilon_{\rm qw}$ is a free parameter which value is 320 as reported by F20. Note that this AGN-driven wind scaling shows results consistent with both hydrodynamical cosmological simulations \citep{Brennan18} and observational findings \citep{Fiore17}. 

\begin{comment}
\citet{Menci19} presents two-dimensional models to calculates the expansion of AGN-driven shocks in a galaxy disc with an exponential gas profile. It considers the outflow expansion in different directions relative to the disc's plane and determines the total mass outflow rate based on the global properties of the host galaxy and the luminosity of the central AGNs. In their work, the fraction $f_{\rm qw}$ of cold gas ejected can be expressed as a function of the bolometric luminosity ($L_{\rm bol}$), the total gas mass ($M_{\rm gas}$), and the virial velocity of the parent dark matter halo ($V_{\rm vir}$):
\begin{equation}\label{eq:outflow}
    \dot{M}_{\rm qw} =  f_{\rm qw} (L_{\rm bol}, M_{\rm gas}, V_{\rm vir}).
\end{equation}
\end{comment}

\subsection{Building the merger trees}
\label{sec:trees}

The standard \gaea runs have been defined on DMH merger trees extracted from the Millennium Simulation suite \citep{Springel05}. The merger tree format adopted in \gaea is organized as follows. 
The starting point is a temporal sequence of {\em snapshots}, where the position and velocity of all particles at a given time is provided.
First, DM halos are identified using a classical Friends-of-Friends (FoF) algorithm.
%which creates a catalogue of virialised halos. 
Then, a second algorithm (SUBFIND) identifies subhalos in each FoF group. Finally, the merger trees are built by identifying unique descendants for all 
%bound 
subhalos (see \citet{Springel05} for details).  For each identified halo or subhalo, pointers and physical quantities are stored at the same redshifts of the snapshots. These merger trees are then provided to {\sc gaea} as a skeleton for the galaxy formation model.
%any given cosmic time or redshift. This collection of data output at each time step is referred to as a \textit{snapshot}. With access to the complete set of snapshots, one can reconstruct the merging history of all halos. 
%At this point, the \gaea model initializes synthetic galaxies based on the host halo properties and follows their evolution throughout the cosmic time as specified by the fixed time grid.

%Following subhalos is a demanding task for an N-body simulation at the relatively low resolution of the Millennium simulation, so their destruction time (defined as the first snapshot in which the subhalo is not found by the halo finder) is not a faithful estimator of the time at which its a satellite galaxy is expected to merge with the central one.
At the resolution of the Millennium simulation, dark matter subhalos \textit{disappear}, i.e. cannot be identified anymore as distinct subhalos, at typically large distances from the halo centre, when the merger is likely incomplete. Since the baryons are more centrally concentrated than dark matter, this time is presumably underestimating the time at which the hosted satellite galaxy is expected to merge with the central galaxy.
In \gaea, this numerical effect is mitigated by modeling orphan galaxies: when a subhalo is lost, its satellite galaxy is assigned a residual merger time, estimated using dynamical friction arguments, and its evolution is followed until it merges with the central galaxy of the host halo. The position and velocities of orphan galaxies are obtained by following the most bound particle of the disrupted subhalo.

%Full N-body simulations (e.g., the Millennium Simulation) utilize specific halo finders to identify bound subhalos. However, these algorithms %experience some difficulties 
%suffer from numerical issues
%when resolving halos close to the resolution limit. In particular, tidal stripping moves matter out of the bounded regions of 
%stripping of dark matter frequently drags matter 
%out of the gravitational potential well of the 
%subhalos, decreasing their mass. At this point the 
%halo finder algorithm does not recognise a previously existing halo which 
%subhalo disappears from the simulation as its mass drops below the DM resolution. Hence, the net effect is to lose 
%otherwise still existing 
%subhalos that would still be resolved at much higher resolution.
%which may host satellite galaxies. 
%are likely to host baryonic structures. 
%is taken into account in order to properly 
%estimate 
%follow the fate of satellite galaxies.
%residing in subhalos, i.e., \textit{satellites}. 
%As a subhalo is lost, 
%not identified anymore, 
%its satellite galaxy is tracked
%living at its center 
%keeps existing independently from the hosting halo, becoming a so-called \textit{orphan} galaxy (to distinguish them from satellites whose substructure is still resolved). 
%Its trajectory until the final merger event is estimated based on dynamical friction arguments, i.e., the orphan is associated with a residual merging time once the host halo is lost in the dark matter-only simulation \citep{DeLucia_Blaizot07}. 

The formalism and the structure of the \pinocchio merger trees (see \S\ref{subsec:pinocchio}) are substantially different compared to the numerical ones. The outputs are provided in the form of merging histories and catalogs of dark matter halos. The merger histories offer a continuous time sampling that uniquely determines the evolution of every single halo along the merger tree, providing the mass, the redshift of first appearance, the merging redshift (if any), the halo ID and other useful pointers. The catalogs supply, according to a time grid, the information about mass, position, velocity and halo ID for all halos. Thus, in order to use \pinocchio outputs as \gaea inputs, it is necessary to adapt the merger trees to a different format, and add quantities that are used by \gaea and not available in \pinocchio as detailed in the following. 

As a first step, for the existing halos, at any given time we linearly interpolate in redshift the position, velocity and mass between the two closest \pinocchio catalogs. In fact, both \pinocchio and \gaea do not require running on a shared, pre-defined time grid. Depending on the specific scientific application, without re-running \pinocchio we can interpolate the physical quantities available in the catalogs according to any given snapshot list. The accuracy of such interpolation is assured by the fact that \pinocchio halo positions are predicted with LPT. This allows us to run \pinocchio once and then freely choose the snapshot list as input to \gaea afterwards. For the purpose of this study, given the relatively small volume, we set twice as many snapshots as in the Millennium, equally spaced in the logarithm of the scale factor $a$ from redshift 127 to 0. This guarantees the self-consistency with the \gaea predictions based on the Millennium merger trees and to concurrently increase the time sampling for a better time resolution. At redshifts above $\sim 4$ we also test a time sampling with 4 times the number of snapshots with respect to the Millennium one and we verify that the predictions are stable and robust as one decreases the integration time step in the semi-analytic model.

Secondly, one physical quantity required by \gaea, but unavailable in \pinocchio, is the maximum rotational velocity of the halo $V_{\rm max}$. In dark matter N-body simulations this quantity is estimated as:
\begin{equation}\label{eq:v_max}
    V_{\rm max} = max\Big(\sqrt{\frac{GM(r)}{r}}\Big).
\end{equation}
The radius $r$ runs over all the particles bound to the halo. In \pinocchio merger trees, such information is not directly available. We utilise instead:
\begin{equation}\label{eq:v_circ}
    V_{\rm max} = f(z) V_{\rm circ},
\end{equation}
where a dimensionless factor $f(z)$ (of the order of unity), which is a function of redshift, multiplies the circular velocity of the halo which can be estimated as:
\begin{equation}
    V_{\rm circ} = \sqrt[3]{10 G H(z) M},
\end{equation}
where $H(z)$ is the Hubble parameter and $M$ the mass of the halo defined and calibrated against FoF halo masses \citep{Monaco02}. 
%For subhalos, their mass is not further used in the F20 version of \gaea to evolve the galaxy during this stage, and we keep their value at the accretion time constant until the final galaxy merger. 

Finally, it is important to note that in \pinocchio, when two halos merge (referred to as the \textit{accretion} time), the smaller one gives its mass to the larger one, and is \textit{removed} from the halo list, meaning that it is not further updated. This means we cannot keep track of these halos in the simulation once they become subhalos. Therefore, we model the presence of subhalos by assigning a survival time to each halo that is lost following a merger. This approach integrates the merger tree with information on its subhalo, as detailed by \citet{Berner22} (see next section). 

% By assuming physically motivated recipes for the missing information, one can estimate the required quantity. Since in \pinocchio a halo disappears from the merger history of the tree as soon as it merged with a more massive one, we need to implement a treatment of substructure belonging to a main halo in terms of spatial and physical properties of subhalos, which are instead assumed to co-evolve within the same cosmic structure. 

\subsubsection{Modeling subhalos}\label{sec:subhalos}
In order to cast the above requirements into the \pinocchio-based approach, we have written a \textit{translation} code that models the presence of subhalos and adds physical information to properly mimic and output Millennium-like DMH merger trees. We use results derived from both theoretically and numerically predicted quantities to fill the required physical information and add the treatment of subhalos. In particular, we assign a population of subhalos within each halo by: 1) assuming a spatial distribution for subhalos following a Navarro-Frenk-White (NFW) density profile \citep{Navarro97}; 2) implementing a statistically derived distribution for the angular momentum of subhalos based on the orbital binding energies observed in numerical simulations \citep{Zentner05, Birrer14}; and 3) assigning a subhalo survival time since accretion \citep{Berner22, Boylan08}. Note that the NFW profile is not accurate for subhalos in the inner regions of the DMHs, but describes well the observed spatial distribution of galaxies \citep[e.g.,][]{Gao04}. We opt to sample the position of subhalos from the NFW distribution at each output time, which means there is no correlation in their spatial position between two consecutive snapshots. In fact, subhalo orbits and positions do not affect the physical treatment of galaxies in the F20 version of the \gaea model. 

We define a total merging time of each subhalo since the merging of the two halos (accretion time) as the sum of the survival time of dark matter subhalos, derived from simulations, and the residual merging time of their galaxies at the time subhalos disappear. This is supported by the fact that in our runs of \gaea, satellite and orphan galaxies are equally treated. 
%Therefore, we assume that the subhalos survive until the galaxy merger while not affecting the evolutionary history of galaxies. As a matter of fact, we circumvent the issue of having orphan galaxies by extending the lifetime of the satellites. This requires modeling the satellite and the orphan phases into the total merging time of subhalos. This is defined as the time from when the two halos merge (accretion time) to the time when the smaller, merging galaxy is added to the central one. 
Our approach takes into account the two phases separately. We initially adopt a halo survival time to shape the first part of the total merging time, practically mimicking the satellite phase $t_{\text{sat}}$. Given the merging subhalo mass $m_{\rm sub}$, the main halo mass $M_{\rm main}$ and the orbital circularity $\eta$ at merger, we assume \citep[with reference to Eq.~(2.6) from ][]{Berner22}:
\begin{equation} \label{eq:t_sat}
    t_{\text{sat}} = A(D) \tau_{\text{dyn}} \frac{\Big(\frac{M_{\text{main}}}{m_{\text{sub}}}\Big)^{b(D)}}{\log\Big(1 + \frac{M_{\text{main}}}{m_{\text{sub}}}\Big)} \exp(c\eta), 
\end{equation}
where  $A$ and $b$ are functions of the linear growth factor $D(z)$ and $c$ is a free parameter. The timescale $\tau_{\rm dyn}$ is the dynamical time at the virial radius, which is assumed to depend on the Hubble constant $H$ only and its value is of the order of $\sim0.1 H^{-1}$.

Next, we add a second time calibrated by estimating the residual merging time of orphan galaxies assigned by \gaea as a function of the halo masses at the time the subhalos disappear in the Millennium simulation. This was originally derived from the classical dynamical friction timescale \citep{Chandrasekhar43, 1987gady.book.....B}, adapted to be applied in SAMs \citep[e.g., ][]{DeLucia_Blaizot07,Boylan08, DeLucia10}, which requires inputs of the position of the subhalo at the time it is lost in the simulation. Since this information is not available in \pinocchio, we assume a dependence only on the mass ratio between the merging halo ($m_{\rm sub}$) and the main halo ($M_{\rm main}$) at the accretion time. We also attempt to factor out the dependence on the redshift by directly fitting the ratio between the extra orphan time $t_{\text{orph}}$ and the age of the Universe, $t_{\text{age}}$, at the accretion time. Based on the above considerations, we derive a formula for the orphan time, given by:
\begin{equation} \label{eq:t_orph}
    t_{\text{orph}} = t_{\text{age}} k \frac{\Big(\frac{m_{\rm sub}}{M_{\rm main}}\Big)^{\alpha}}{\log\Big(1 + \frac{m_{\rm sub}}{M_{\rm main}}\Big)^{\beta}},
\end{equation}
where $k$, $\alpha$ and $\beta$ are free parameters (see the next section for the calibrated values).

Finally, the total merging time of the satellite galaxy is the sum of the estimated halo survival time as in Eq.~\eqref{eq:t_sat} and the modelled extra time for orphan galaxies given by Eq.~\eqref{eq:t_orph} (see Appendix~\ref{appendix:time}). Note that while the individual contributions from these time scales vary depending on the mass resolution of the DMH trees, their combined total remains resolution-independent. This is because the total merging time is determined by the halo mass ratio and the accretion redshift. 
Hence, directly adding the total merging time to newly identified subhalos ensures that the construction of the \pinocchio merger trees does not depend on the DM mass resolution. 

\subsection{Calibrating the trees}
\label{sec:calibration}

As discussed in \S\ref{sec:trees}, our reference \gaea realization is run on merger trees extracted from the Millennium Simulation. This is a numerical realization of a cosmological volume of side 500 $h^{-1} \text{Mpc}$ assuming the WMAP1 lambda cold dark matter cosmology \citep[$\Omega_\Lambda = 0.75$, $\Omega_m = 0.25$, $\Omega_b = 0.045$, $n = 1$, $\sigma_8 = 0.9$, and $H_0 = 73 \, \text{km s}^{-1} \, \text{Mpc}^{-1}$, ][]{Springel05}. We run a \pinocchio box that matches the Millennium one in terms of particle mass resolution, volume and cosmological parameters. This run has been used to calibrate the time scales discussed in the previous section.

\begin{figure}
	\includegraphics[width=\columnwidth]{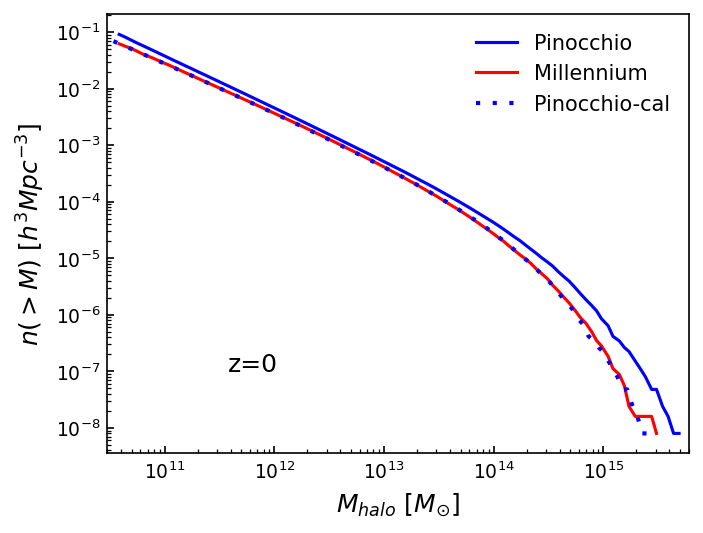}
    \caption{Halo Mass Functions (HMF) at redshift $z=0$ from the \pinocchio Millennium-like box (solid blue line) and the Millennium simulation (solid red line). Due to the different definition of the halo mass, at $z=0$ the halo masses in \pinocchio are moderately higher compared to the Millennium halos. We calibrate the \pinocchio HMF by using a polynomial fit of mass difference at a given number density that is a function of the redshift (dotted blue line).}
    \label{fig:halo_mf}
\end{figure}

The different definition of halo mass between the two simulations results in DM halo masses which are on average a factor of $\sim$2-3 more massive in \pinocchio with respect to the Millennium (see Fig.~\ref{fig:halo_mf}). This difference requires us to calibrate, at each snapshot, the \pinocchio Halo Mass Function (HMF) to match the Millennium one. To do so, we bin the range of cumulative number density $n$ between the maximum halo mass for which we count at least ten objects and the minimum halo mass corresponding to the resolution of the simulation ($\sim5 \times 10^{8}M_\odot$). In such an interval, we fit both the \pinocchio and Millennium HMFs and we retrieve their ratio in halo mass for any given number density. Then, we multiply the \pinocchio halo mass by the correspondent correction factor. This operation provides halos with the same mass distribution of the Millennium simulation, thus avoiding introducing systematic bias in subsequent steps. With a similar strategy, we calibrate the distribution of the maximum rotational velocity of the halo, $V_{\rm max}$, introducing a correction factor in order to match the distribution extracted from the Millennium main halos as a function of redshift (see Eq.~\eqref{eq:v_circ}). Additionally, the process of the assembly of the merger trees involves parameters we need to adjust, specifically the ones associated with the estimate of the merging time.  For this scope, we run the \gaea semi-analytic model on the \textit{translated} merger trees generated from the \pinocchio box as described above. Then we aim at calibrating the parameters involved in the construction of the merging times $t_{\text{sat}}$ and $t_{\text{orph}}$ by reproducing the galaxy stellar mass function (GSMF) at redshift $z=0$. In fact, the merging history of galaxies plays a crucial role in the build-up of their stellar and gas mass content, as well as shaping the cosmic star formation rate density (SFRD) and AGN activity across the age of the Universe. % (e.g., \citet{Heckman14, Somerville15, Pillepich18, Dave19}).

\begin{comment}
Indeed, the shape of the GSMF has been studied extensively: the flattening at low masses (below the knee) is thought to be mainly driven by the self-regulation of the star formation process by means of gas heating via SNa feedback, while the exponential cutoff at the massive end is believed to result from AGN feedback efficiently reheating cold gas otherwise available for star formation . 
Both phenomena depend on the merging history of the galaxy, as bursts of star formation and AGN activity are often modelled as triggered by merger events (\citet{Fontanot20}). 
Therefore, the position and the normalization of the knee in the GSMF offer crucial insights. Specifically, this information is instrumental in addressing the galaxy population where the majority of the stellar mass has been assembled in, throughout the age of the Universe.   
\end{comment}

\begin{figure}
	\includegraphics[width=\columnwidth]{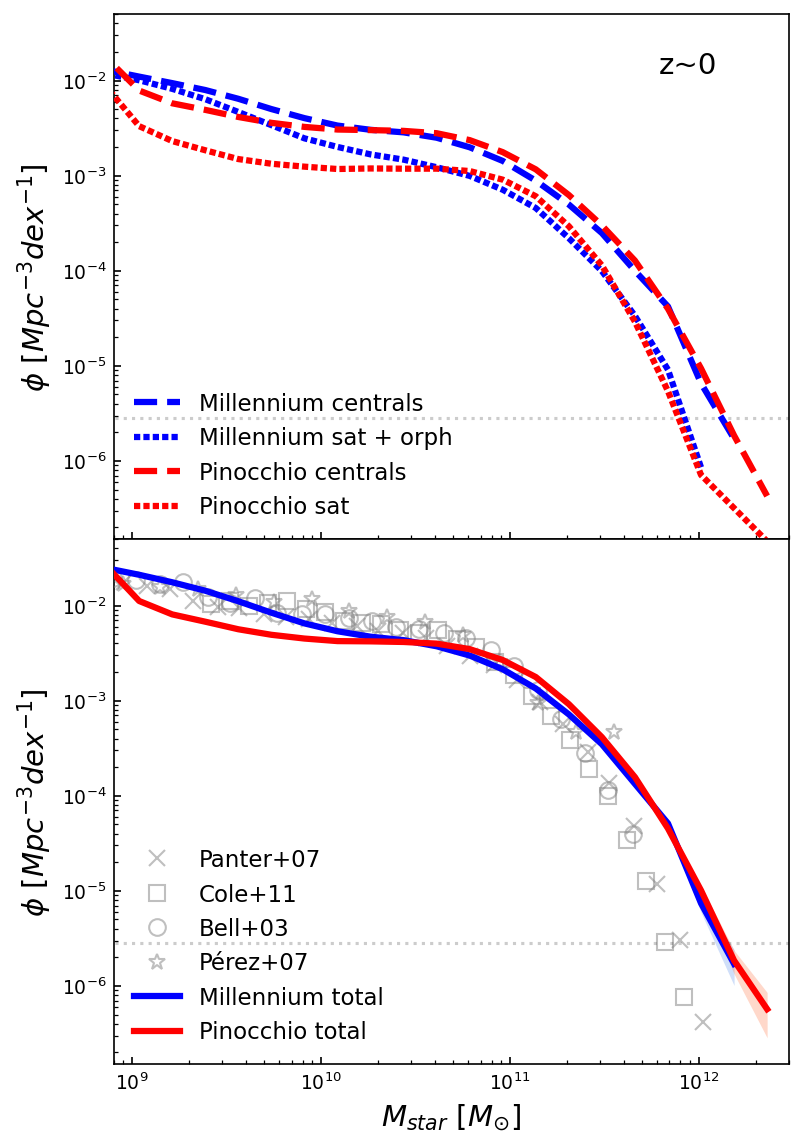}
    \caption{\textit{Upper panel:} GSMF at redshift z=0 from the Millennium-like box. Dashed (dotted) lines indicate the contribution of the central (satellite and orphan) galaxies, respectively. While central galaxies show remarkable agreement, Pinocchio's low mass satellites differ by a factor of $\sim 4$ in number density with respect to the Millennium ones. \textit{Lower panel:} Total GSMF at $z=0$ from the Millennium-like box. The symbols show observed data points from \citet{Panter07, Cole01, Bell03, Perez08}. This plot shows the goodness of the calibration, especially above $10^{10} M_{\odot}$.}
    \label{fig:smf_z0_mill}
\end{figure}

We explore different combinations of the satellite $t_{\text{sat}}$ and orphan $t_{\text{orph}}$ time scales in order to reproduce the normalization and the position of the knee of the GSMF at $z\sim0$. In particular the total merging time of the satellite galaxies impacts on the total number density in terms of their relative contribution to the GSMF. Therefore, we calibrate the parameters in Eqs.~\eqref{eq:t_sat}) and \eqref{eq:t_orph} by matching at the same time the GSMF of central and satellite galaxies obtaining the following values:
\begin{equation}\label{eq:params}
    \begin{aligned}
        A &= \begin{cases} 
        0.39 & \text{for } D(z) > 0.8 \\
        0.195 + \frac{0.195}{0.2}(D(z)-0.6) & \text{for } 0.8 \geq D(z) \geq 0.6 \\
         0.195 \left( \frac{D(z)}{0.6} \right)^2 & \text{for } D(z) < 0.6 
        \end{cases}\\
        b &= 1.015 \cdot D(z)\\
        c &= 1.3\\
        k &= 0.3\\
        \alpha &= 0.94\\
        \beta &= 1.7.
    \end{aligned}
\end{equation}

Fig.~\ref{fig:smf_z0_mill} shows the result of our best estimate of the GSMF from both populations of galaxies (upper panel). The GSMF of centrals using the \pinocchio merger trees agrees well with the Millennium-based predictions for stellar masses above few times $10^{10} M_{\odot}$, throughout the knee until the exponential cut-off at the high-mass end. On the other hand, satellites show a substantial difference (a factor of up to $\sim 4$) at the low mass end of the GSMF, i.e., at masses below $\sim 10^{10} M_{\odot}$, even though the total gap in the bottom panel is dominated by the central galaxies. We argue that this difference is due to the intrinsic differences in the features of the \pinocchio and Millennium simulations in the definition and construction of merger events (orbit crossing vs FoF).
%ii) the definition of halo mass in the two methods (i.e., being slightly more massive in \pinocchio). 
As a consequence, this effect causes a dearth of small mass halos as one approaches $\sim$ few $10^{9} M_{\odot}$, explaining the gap in the number density of galaxies.  However, if one compares the total GSMF combining all types of galaxies (bottom panel), the difference in the low-mass end reduces to a factor of $\sim 2$ or less.
\begin{comment}
Moreover, it has been shown that N-body simulations tend to overpredict the number of galaxies in this small mass regime (see for example \citet{Fontanot09}).The overabundance of massive galaxies of our predicted GSMFs with respect to the observed trend can be explained with a probabilistic argument. As mentioned at the beginning of this section, the effective volume considered for the calibration is actually a fraction of the original box, large enough to host a cluster of galaxies. This allows us to first considerably speed up the computation and second to sample the widest possible range of galaxy masses. Within such an approach, the price to pay is that one should not expect the predicted GSMF to reproduce the massive tail observed in the data. In fact, the massive objects will naturally be overabundant since they are generated by the larger modes produced by the initial conditions in the original box, otherwise inaccessible in smaller volumes.
\end{comment}
We also verify that at earlier epochs, when the fraction of satellite galaxies is lower, the impact of the merging time is marginal. Our GSMF reproduces the trend obtained from the Millennium-based predictions with a better agreement compared to $z \sim 0$.
Bearing in mind these caveats, we consider the calibration resulting in the GSMF depicted in Fig.~\ref{fig:smf_z0_mill} as being sufficiently accurate for the scope of this work and we opt not to calibrate the \gaea physical parameters keeping the values proposed by F20 (see also Appendix~\ref{appendix:gsmf}). 

\subsection{Seeding SMBHs}
\label{sec:seeding}

%In the early universe, 

\begin{figure}
\centering
	\includegraphics[width=0.7\columnwidth]{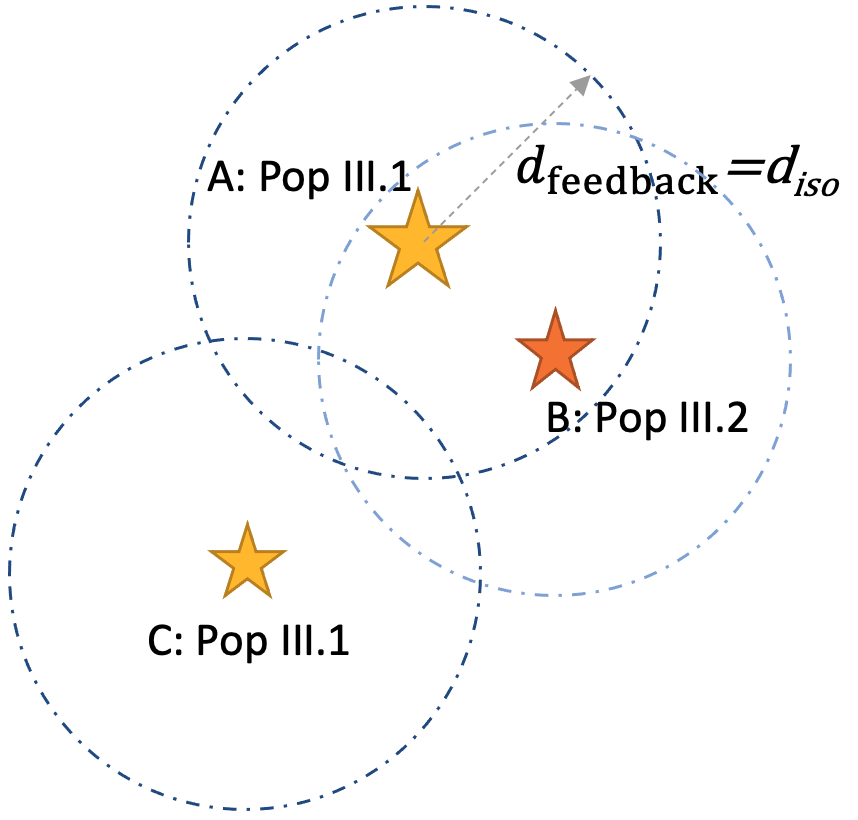}
    \caption{A simplified illustration of the Pop III.1 SMBH seeding scheme (see text) showing the conditions for a star to be isolated enough to be considered as a Pop III.1 star. Adapted from \citet{Singh23}.}
    \label{fig:popIII}
\end{figure}

The primary goal of our paper is to investigate the impact of the Pop III.1 SMBH seeding scheme on galaxy and SMBH properties. This seeding scheme is illustrated in Fig.~\ref{fig:popIII} 
%to identify which halos give rise to Pop III.1 stars and subsequently to SMBHs 
\citep[for a more detailed introduction of the seeding we refer to ][]{Banik19, Singh23}. In the figure, three stars, denoted as A, B, and C, reside in different halos. Among them, only A and C evolve into Pop III.1 supermassive protostars, while B is designated as a lower mass Pop III.2 star. This classification depends on their physical separation from other sources at the time of their formation.

Star A, forming first, exerts its influence within a sphere of radius equal to $d_{\rm feedback}$, primarily driven by radiative feedback. Since this star originates in a pristine primordial gas environment, devoid of any influence from neighboring stars, it falls under the category of a Pop III.1 star. Star B, on the other hand, forms at a distance less than $d_{\rm feedback}$ from star A, subjecting it to the effects of feedback. Consequently, it is classified as a Pop III.2 star (or even a Pop II star if the gas cloud it forms from is chemically enriched). Finally, star C forms beyond the regions influenced by feedback from stars A and B. As a result, it is are designated as another Pop III.1 source.

The physical mechanism that may allow SMBH formation is the impact of dark matter annihilation heating on the structure of the Pop III.1 protostar \citep{Spolyar08,Natarajan09,Rindler-Daller15}. The expected effect is to keep the protostar relatively large and cool, thus emitting a weak flux of ionising radiation. Efficient collapse of the baryonic content of the minihalo to the Pop III.1 protostar, i.e., yielding a mass of $\sim 10^5\:M_\odot$, is assumed to occur. After a few Myr of evolution, this source is then expected to form a SMBH of similar mass. Thus the initial mass of the SMBH seeds that we implement in the Pop III.1 scheme is $M_{\rm BH}=10^5\:M_\odot$.
%, and after few Myr they are expected to evolve into an SMBH with mass of the order:
%\begin{equation}
%    M_{\text{BH}} \simeq 10^5 M_{\odot}, 
%\end{equation}
%according to the predictions of the Pop III.1 model powered by dark matter self annihilation. This values is assumed to be the BH seed mass implemented in our framework.

The main parameter of the Pop III.1 model is the isolation distance, $d_{\rm iso}$, which is required for a minihalo to form a Pop III.1 source. \citet{Banik19} and \citet{Singh23} have shown that values of $d_{\rm iso}\simeq 50$ to $100\:$kpc (proper distance) are sufficient to yield overall numbers of SMBHs consistent with estimates of local number densities at $z=0$. We will thus consider values of $d_{\rm iso} = 50, 75, 100\:$kpc in our modeling.

%In this model, the feedback distance is referred to as the isolation distance $d_{\rm iso}$. In essence, for a star to be classified as a Pop III.1 star, it must form in an environment where no previously formed halos exist within the sphere of radius $d_{\rm iso}$. We treat $d_{\rm iso}$ as a free parameter in our scheme, adjusting it to align with the observed number density of SMBHs and galaxy scaling relations in the local Universe.

We also consider two other seeding models. The first of these is the BH seeding scheme that is implemented within \gaea and described in detail in \citet{Xie17}. Recall that this scheme has been tested at the Millennium simulation resolution and is intended to be a sub-grid model that is not attempting to mimic any seeding scenario. Whenever a new DMH is resolved in the \pinocchio simulation, we assign it a BH mass (\(M_{\text{BH}}\)) scaled with the parent DMH mass (\(M_{\text{DM}}\)):
\begin{equation}\label{eq:volonteri_seed}
    M_{\text{BH}} = \left(\frac{M_{\text{DM}}}{10^{10} M_{\odot} h^{-1}}\right)^{1.33} \frac{10^{10} M_{\odot} h^{-1}}{ 3 \times 10^{6}},
\end{equation}
where the index of the relation is derived from \citet{Volonteri11}. We refer to this scheme in which all halos are seeded as ALS (All Light Seed). Note that this is an \textit{ad-hoc} model that mimics the formation of seeds from Pop III stars and estimates their evolution by accretion before the halo is resolved. Given the relatively high mass resolution of our \pinocchio box (see \S\ref{subsec:pinocchio}), applying Eq.~\eqref{eq:volonteri_seed} to the \(M_{\text{DM}}\) mass distribution results in BH seed masses of the order of \(\sim 10^{1-2} M_{\odot}\). This mechanism effectively depicts a scenario where every single halo is seeded with a Pop III remnant BH and can be considered an implementation of the standard light-seed Pop III model described in \S\ref{sec:intro}.  Note that this seeding scheme has been tested previously only at the resolution of the order of the Millennium Simulation \citep{Xie17}. The \gaea version presented in F20 and used in the model calibration described in \S\ref{sec:calibration} is based on this latter seeding scheme.  

%Finally, we also consider model predictions for the HMT seeding scheme as implemented in the Illustris-TNG simulations \citep{Bhowmick22}. In particular, every DM halo exceeding the mass of $7 \times 10^{10} M_{\odot}$ is seeded with a BH of mass $1.4 \times 10^{5} M_{\odot}$. Again, such a model is not related to any physical mechanism of SMBH formation.

Finally, for continuity with Paper II we also consider the HMT seeding scheme, where all halos with mass $>M_{\rm thr}=7.1\times 10^{10}\ M_\odot$ are seeded with a $1.4\times10^5\ M_{\odot}$ BH. This is the typical seeding scheme used in hydro simulations \citep[e.g.,][]{Vogelsberger14}, however the value of the threshold mass is determined by the mass resolution of simulations able to resolve galaxies in a cosmological box. The box used to obtain our results has a much higher resolution, so this scheme can be considered as a toy model for a seeding scheme where BH seeds appear later in the evolution of the Universe.

\section{Results}
\label{sec:results}

Here we present the predicted properties of the galaxy populations based on DMH merger trees extracted from the $40$ Mpc $h^{-1}$ \pinocchio box described in \S\ref{subsec:pinocchio} and \textit{translated} as explained in \S\ref{sec:trees}. We evaluate the Pop III.1 seeding model for three cases of $d_{\rm iso}=50$, $75$, and $100$ kpc (see \S\ref{sec:seeding}), as well as for the ALS and HMT seeding schemes. Unless otherwise specified, we apply the accretion scheme onto the SMBH described in \S\ref{subsec:accretion}.
%and the comparison between the populations of seeded and unseeded galaxies. 
%jct - let's present some results first.
%We defer a study of the high redshift observational implications of the Pop III.1 seeding model, such as the UV luminosity function, black hole masses and bolometric luminosities to an accompanying paper. 
%jct - I do not understand this:
%It is worth noting that since the calibration of the model, see \S\ref{sec:calibration}, has been done by reproducing the evolution of GSMF across cosmic time, the predicted properties of SMBHs and galaxies in our simulation constitute real predictions of the model. 
Throughout this work, the three cases of isolation distance $d_{\rm iso}=50$, $75$ and $100$ kpc are depicted in orange, red and blue, respectively, while the ALS and HMT models are shown in magenta and green, respectively.

%jct - already described.
%The choice of these values is motivated by the results obtained Paper II where this range of $d_{\rm iso}$ reproduces the local observational estimates on the SMBH number density.
\subsection{Occupation fractions}
\label{sec:occ_fract}

\begin{figure*}
	\includegraphics[width=\textwidth]{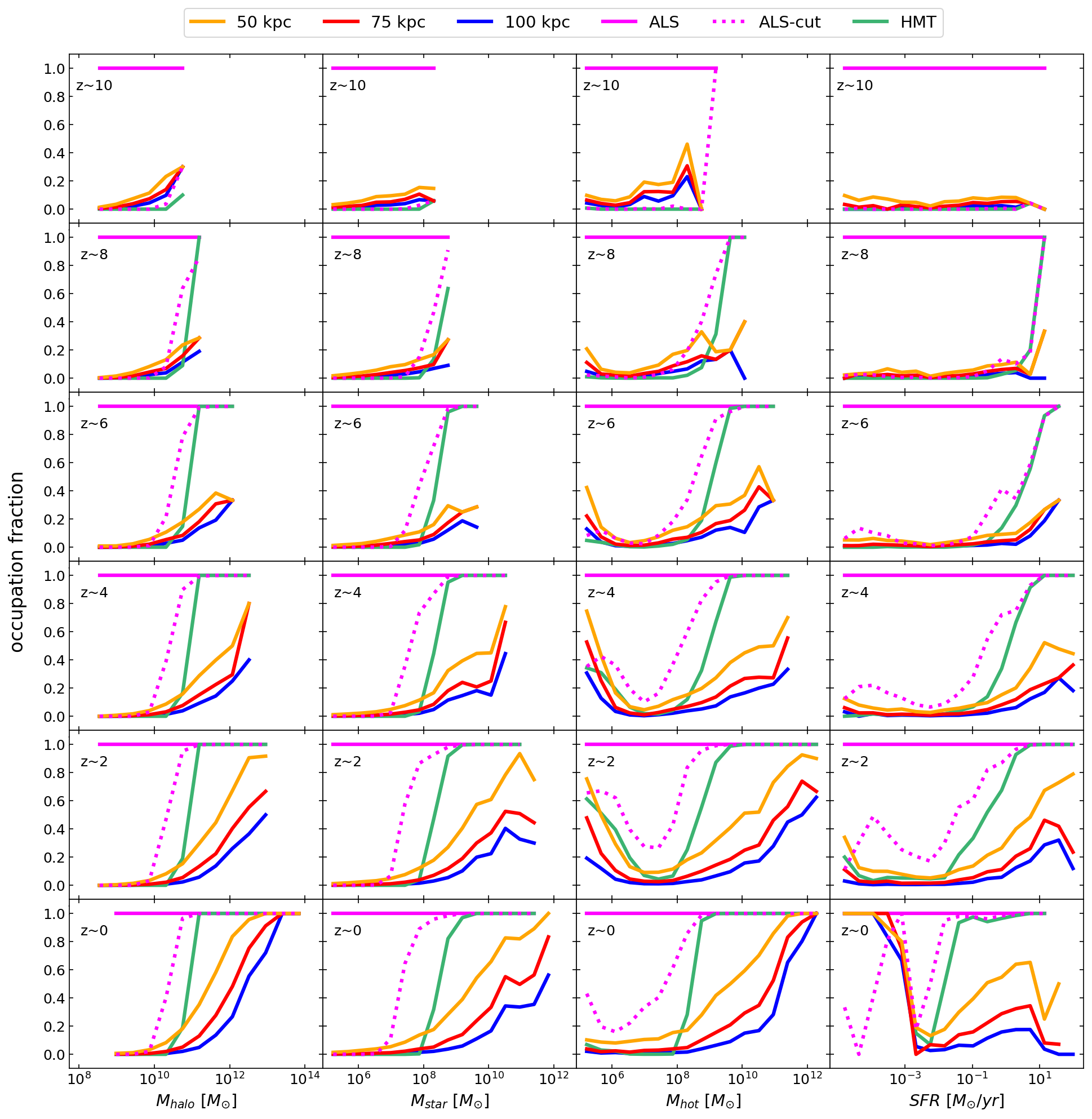}
    \caption{Occupation fraction of seeded galaxies as a function of different galaxy properties for several redshifts. From left to right: halo mass, stellar mass, hot gas mass and star formation rate (SFR). The 6 lines show different seeding mechanisms.}
    \label{fig:occ_fract}
\end{figure*}

\begin{figure}
\centering
	\includegraphics[width=\columnwidth]{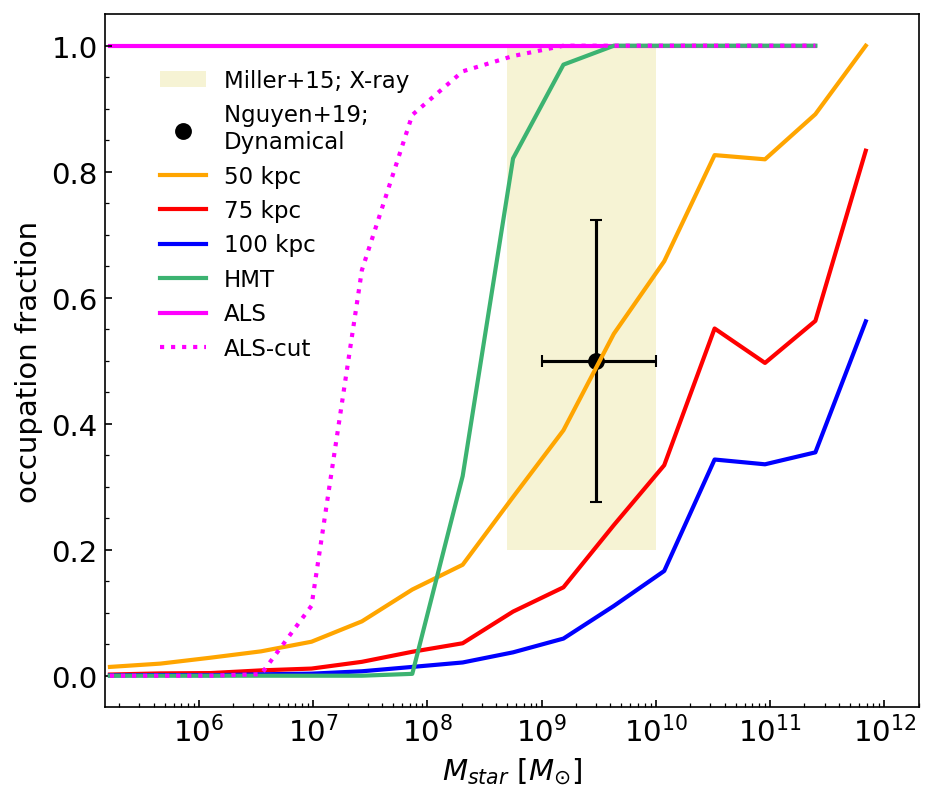}
    \caption{Occupation fraction of seeded galaxies as a function of the stellar mass at redshift zero. The 6 lines show different seeding mechanisms. In the range of stellar masses between few in $10^8$ and $10^{10} M_{\odot}$, we compare against different observational constraints for the occupation fraction. X-ray sources detected in local surveys pose some lower limit to the occupation fraction as presented by \citet{Miller15} drawn as the yellow shaded region. In black we show the more stringent constraints as reported by \citet{Nguyen19} obtained from dynamical findings in a small sample of nearby galaxies. %while the gray shaded area stands as a reference for an upper limit assuming all nuclear star clusters in Virgo hosts a BH seed (\citet{Sanchez-Janssen19}).
    }
    \label{fig:occ_fract_zoom}
\end{figure}

\begin{comment}
\begin{figure*}    
	\includegraphics[width=\textwidth]{pdf_comparison3.png}
    \caption{Probability density function of seeded (solid lines) and non-seeded (dotted lines) galaxies as a function of different galaxy properties for several redshifts. The three lines show different isolation distances.}
    \label{fig:pdf_comp}
\end{figure*}
\end{comment}

%PaperII and \S\ref{sec:seeding} introduce the conditions under which a halo is seeded with a $10^5 M_{\odot}$ SMBH according to the Pop III.1 model. This seeding scheme impacts the relative number of seeded galaxies as a function of halo mass and  we aim to understand its potential implications in terms of other galaxy properties. 

We examine the cosmic evolution of the fraction of galaxies that are seeded with SMBHs, i.e., the \textit{SMBH occupation fraction}, including as a function of various physical properties.
In Fig.~\ref{fig:occ_fract} we show the evolution of the SMBH occupation fraction, i.e., the fraction of seeded galaxies, as a function of various galaxy properties from redshift $z\sim10$ down to $z\sim0$. 
%for the 3 fiducial values of $d_{\rm iso}$ plus the HMT model. 
Different columns indicate different galaxy quantities, from left to right: halo mass, stellar mass, hot gas mass, and star formation rate (SFR). 
%For reference we illustrate 
The ALS seeding scheme is by construction always identical to unity. For a more meaningful comparison in terms of occupation fractions of SMBHs, here we also consider the ALS case applying a cut in BH mass above $10^{5}\:M_\odot$ (hereafter depicted with the label ALS-cut).

At all epochs, the SMBH occupation fractions as a function of halo mass, stellar mass and hot gas mass have similar behaviors across Pop III.1 seeding models. As the isolation distance decreases, the occupation fraction rises. SMBHs tend to reside in the more massive systems, as the Pop III.1 model places them in the first locally formed mini-halos, which are more likely to be the progenitors of the largest structures later on. While the high-mass end occupation fraction as a function of halo mass reaches unity for all models at redshift zero, this is not the case for the occupation fraction as a function of stellar mass, i.e., only $\sim 60\%$ of galaxies with highest level of stellar mass ($\sim 10^{12}\:M_\odot$) are seeded in the $d_{\rm iso}=100\:$kpc case, but rising to about 100\% in the $d_{\rm iso}=50\:$kpc case. The various Pop III.1 cases show occupation fractions that rise gradually for $M_{\rm star}\gtrsim 10^7\:M_\odot$ to $10^9\:M_\odot$ as $d_{\rm iso}$ rises from 50 to 100~kpc. On the other hand, the HMT model shows an occupation fraction that rises very steeply to unity for stellar masses $\gtrsim {\rm few}\times 10^9\:M_\odot$ across all epochs. The ALS-cut case presents a similar shape, but shifted downward by one order of magnitude to $10^8\:M_\odot$ by z$\sim$0. This reflects the fact that, when a low mass seeding scenario is applied, the viscous accretion mode soon establishes a BH growth rate proportional to the SFR reproducing the local scaling at the latest epochs ($M_{\rm star}\sim 10^3 M_{\rm BH}$). Therefore, applying a cut in $M_{\rm BH}$ corresponds to impose a sharp threshold for $M_{\rm star}$.

Thus a key difference between the Pop III.1 models and the HMT and ALS seeding schemes is that the Pop III.1 models have much smaller SMBH occupation fractions at relatively high values of $M_{\rm star}$. This implies a population of unseeded galaxies reaching large values of $M_{\rm star}$. Indeed, in the absence of AGN feedback, these systems keep forming stars with relatively high rates at all redshifts, which influences the SMBH occupation fraction versus SFR, discussed below. A related consequence is that the Pop III.1 models have a larger scatter in the $M_{\rm star}-M_{\rm halo}$ relation, discussed in \S\ref{sec:bh_mstar}.

%Therefore, given a fixed $M_{\rm star}$ we observe a larger spread in the corresponding $M_{\rm halo}$ in the Pop III.1 models, thus explaining the different behavior in the occupation fractions.
%larger scatter in the $M_{\rm star}-M_{\rm halo}$ relation, especially at low redshift, including significant populations of unseeded galaxies . This scatter is greatly increased in the Pop III.1 seeding scenarios due to a population of unseeded galaxies reaching larger $M_{\rm star}$ with respect to the HMT and ALS schemes. In fact, in the absence of AGN feedback, these systems keep forming stars with larger rates at all redshifts. Therefore, given a fixed $M_{\rm star}$ we observe a larger spread in the corresponding $M_{\rm halo}$ in the Pop III.1 models, thus explaining the different behavior in the occupation fractions.}}

% which shows a well defined main sequence extending above $M_{\rm star}>5\times10^10\:M_\odot$ for $d_{\rm iso}>50\:$ kpc at $z=0$. 

When considering SMBH occupation fraction as a function of hot gas mass, to the extent that more massive halos and/or stellar components correlate with $M_{\rm hot}$, we see similar trends. However, at intermediate redshifts we see that there is a population of seeded galaxies with very low values of $M_{\rm hot}$, which are attributed to systems in which AGN feedback prevents the build up of hot gas mass.

%However, many seeded galaxies have very low SFRs, i.e., elliptical galaxies, so we see the $z=0$ occupation fraction rise to unity for at low SFRs.

%the stellar and hot gas mass show different values as their total budgets depend on physical phenomena, i.e. star formation, AGN and stellar feedback. 
%jct - I do not understand the next sentence:
%For instance, the cold gas available for star formation is related to the presence of a SMBH, hence sensitive to $d_{\rm iso}$ up to large stellar masses. 

%As the halo and stellar mass share a monotonically increasing behaviour, the hot gas mass depicts a raise at lower masses, especially at earlier times whereas it vanishes at z$\sim$0. 

%This can be explained with the additional information that, even if few low mass galaxies are seeded, AGN feedback heats up and eventually expels part of the already low gas content (via AGN-driven winds) keeping the hot gas mass low. 
%Conversely to non-seeded galaxies, this prevents seeded ones to accumulate hot gas. 

%jct - I can't see the number density.
%After z$\sim$2 the number density of AGNs decreases sharply and by redshift 0 the seeded galaxies have gravitational potential deep enough to make this effect negligible. 

%Galaxies without seed do not experience the same phenomena, effectively building up their hot gas content. 

Fig.~\ref{fig:occ_fract} also shows the evolution of the occupation fractions as a function of SFR. At later epochs, from redshift $\sim$ 4 down to intermediate redshifts (z$\sim$2), one sees how seeded galaxies, mostly residing in more massive galaxies, tend to have larger SFR. At the same time AGN feedback begins to influence the gas in these galaxies, leading to an efficient suppression of star formation in massive systems. This demonstrates the effectiveness of radio-mode AGN feedback in keeping galaxies passive. By redshift z$\sim$0 the fraction of passive galaxies, characterized by SFRs lower than  $10^{-3} M_{\odot}/yr$, reaches SMBH occupation fractions of unity across all models. However, we see that the ALS-cut scheme reaches one about SFR of $10^{-3} M_{\odot}/yr$, but it drops down to small fractions soon after. One should keep in mind that ALS-cut does not track the whole population of seeded galaxies, but only the ones having $M_{\rm BH}>10^5\:M_\odot$. As opposed to the other models, this latter sub sample of galaxies is not exclusive in terms of radio-mode AGN feedback since this also affects galaxies with BHs just below the BH mass cut, contaminating the measurements of the occupation fraction for the SFR as well as for the hot gas.

As a point of comparison with observational data, in Fig.~\ref{fig:occ_fract_zoom} we illustrate constraints for the fraction of observed local galaxies hosting a SMBH in the range of stellar masses between few in $10^8$ and $10^{10} M_{\odot}$. Observationally, dynamical and X-ray measurements based on local galaxies seem to agree that at least $\gtrsim50\%$ of the population in this mass range host a SMBH with mass  $\sim10^{4-6} M_{\odot}$ \citep[see discussion in ][]{Greene20}. In particular, \citet{Nguyen19} show that out of 10 galaxies within a distance of 4 Mpc for which dynamical measurements are available, only 5 SMBHs are detected, inferring a lower limit for the occupation fraction. Additionally, \citet{Miller15} advocate a lower limit of 20\% with a most probable estimate of around 70\% based on the detection of X-ray sources in local galaxies within a comparable stellar mass. 

Fig.~\ref{fig:occ_fract_zoom} illustrates that current observational measurements are not of sufficient precision to discriminate between the different models. If the quality and statistics of observational samples can be increased to reduce the uncertainties, then the presented models could be distinguished. However, it is interesting to note that our simulations present a better agreement in terms of the occupation fraction when $d_{\rm iso}$ approaches the value of 50 kpc, as we allow more halos to be seeded. Conversely, the 75 and 100 kpc cases struggle to produce large enough occupation fractions while the ALS-cut and HMT model tends to create more numerous SMBHs although still in agreement with \citet{Miller15}. On the other hand, the ALS scenario would be difficult to test against current observations as the totality of galaxies in this mass range would host a BH, whose mass would span from stellar seeds up to the SMBH regime.

\subsection{Stellar mass function}
\label{sec:smf}
\begin{figure*}
	\includegraphics[width=\textwidth]{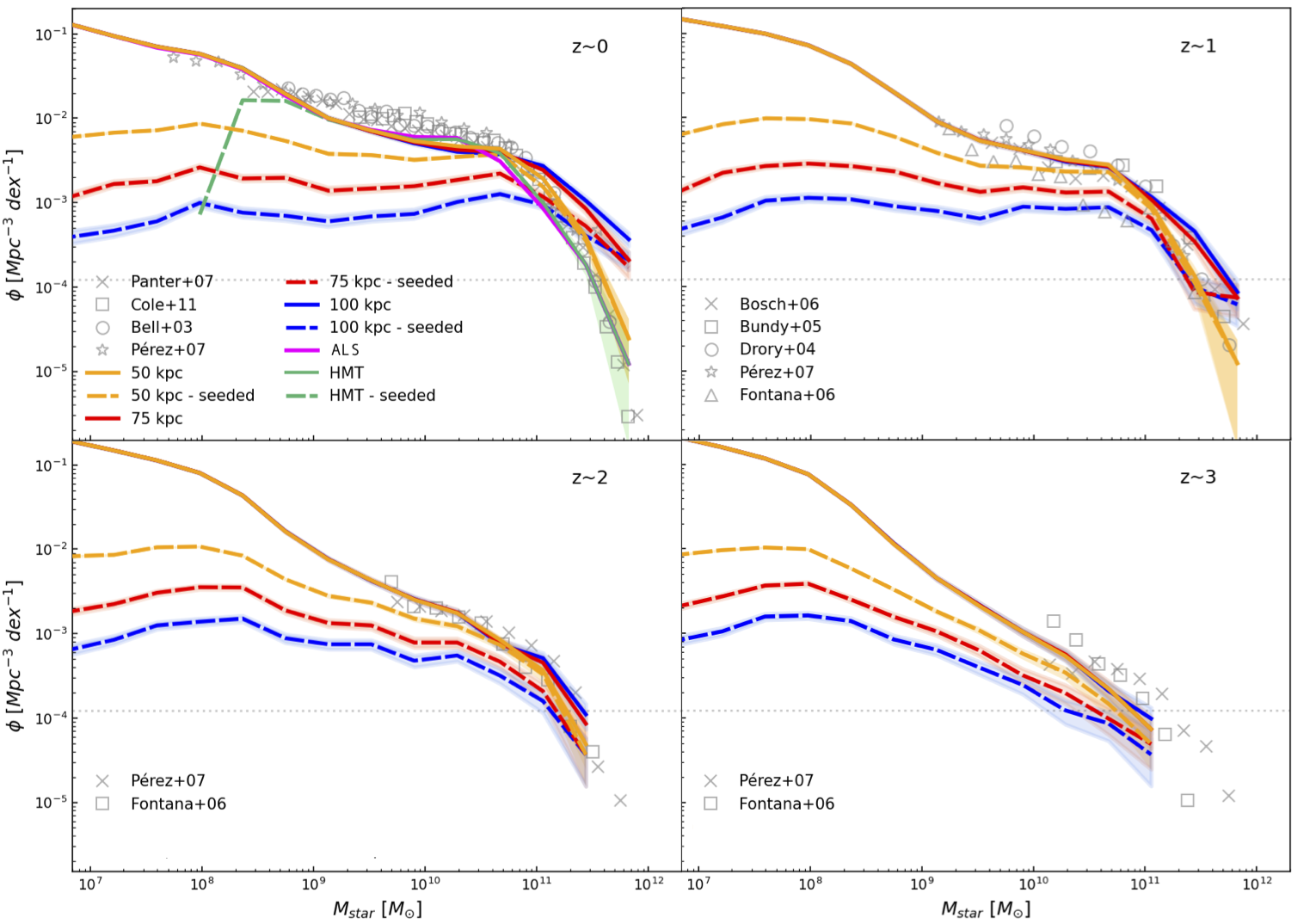}
    \caption{Cosmic evolution of the GSMF from redshift 0 up to 3. Solid (dashed) lines indicate the contribution of all (seeded) galaxies, respectively. The shaded areas depict the Poisson uncertainties in each mass bin. Results for Pop III.1 models with 3 values of isolation distance (in proper distance) are shown, as labelled. Observed data are in gray symbols \citep{Panter07, Cole01, Bell03, Perez08, VanDenBosch08, Bundy05, Drory04, Fontana06}. As a reference, the dotted horizontal line stands for 10 objects per mass bin in the whole volume of the box. The GSMF tells us the isolation distance parameter that better reproduces the observed trend, especially at the massive end. The 50 kpc case agrees with the exponential cut off of the massive population of galaxies from the local relation, where almost every galaxy with a stellar mass higher than \(\sim 3\times10^{10} M_{\odot}\) is assumed to host a SMBH, up to $z\sim 1$. At redshift 0 we compare with the HMT and ALS seeding schemes. These two cases mostly overlap in the graph. While ALS well reproduces the local observational trend by construction, HMT naturally seeds every massive galaxy resulting in efficient quenching of star formation in such systems. We use them as reference models to validate the Pop III.1 seeding at redshift z=0. The green dashed line of the HMT scheme tells us that the fraction of seeded galaxies rises sharply up to one above $\sim 10^{9} M_{\odot}$ as shown from Fig.~\ref{fig:occ_fract_zoom}.}
    \label{fig:smf_z0}
\end{figure*}

In Fig.~\ref{fig:smf_z0} we illustrate the cosmic evolution of the GSMFs evaluated for the Pop III.1 seeding models 
%$d_{\rm iso}$ already introduced in the previous section spanning 
from $z\sim0$ up to 3. For the redshift zero case, we see that below stellar masses of approximately $3\times10^{10} M_{\odot}$ the predicted functions match reasonably well down to low masses of $M_{\rm star}\sim 10^8\:M_\odot$. In this regime, the majority of galaxies are non-seeded, and so there is little impact, i.e., due to AGN feedback, of varying $d_{\rm iso}$.
%as the contribution to the total GSMF from the seeded ones is marginal and no strong dependence on the isolation distance is shown in terms of the total number density. 
%Moreover, the agreement with the data is remarkably good, even though we do observe a deficit of a factor of $\sim 2$ at most for masses about $\sim 10^{10} M_{\odot}$, as previously addressed in \S\ref{sec:calibration}. 
%Note that the merger trees fed into \gaea have been calibrated on a different box with a lower resolution, see Fig.~\ref{fig:smf_z0_mill}, and a recalibration of the model may absorb this difference. 
On the other hand, reducing $d_{\rm iso}$ from 100 to 50~kpc has a dramatic impact at the high-mass end of the GSMF.
%the three cases of $d_{\rm iso}$ present different results in terms of the massive end of the GSMF. 
Specifically, while the 75 and 100 kpc models struggle to reproduce the exponential cutoff of the GSMF above a few~$\times 10^{10}\:M_\odot$, the  
%usually observed above the knee, the 
50 kpc case is able to maintain good agreement with the observational data. This is due to the action of AGN feedback in suppressing the build-up of high stellar mass galaxies.
%the only one able to keep a good agreement within uncertainties to the GSMF obtained from an 
We see that the ALS and HMT models, having similar, saturated occupation fractions at high stellar masses, are also able to reproduce the local high-mass end of the GSMF. It is worth stressing that the agreement with observational constraints cannot be improved by a different parameter calibration, as the lack of a central SMBH in unseeded galaxies prevents the onset of Radio-mode feedback in the first place, and stellar feedback alone is insufficient to reduce the number of massive galaxies while keeping a good agreement at lower masses and using realistic loading factors \citep{WhiteFrenk91}.

Moving to higher redshifts, the differences between seeded and non-seeded galaxies become less visible in terms of average stellar mass distribution, as the SMBH are not big enough to affect the evolution of the host galaxy by means of radio-mode feedback. This also reflects in the GSMF: if we look at the evolution of the GSMF up to redshift $z\gtrsim2$ right at the peak of the cosmic SFR, the different isolation distance scenarios become more similar and all of them tend to agree with the observed data (see higher redshift panels). This makes it more difficult to use such an observable to discern among the seeding criteria, i.e., $d_{\rm iso}$, at $z\gtrsim1$. 

%Note that up to this redshift the massive end of the GSMFs hints that the 50 kpc case for $d_{\rm iso}$ works better in reproducing the exponential decrease of the data. 

%However, the fraction of seeded galaxies in the low mass range can still be used as a test to possibly diagnose and rule out specific seeding mechanisms (Fig.~\ref{fig:occ_fract_zoom}). 

\begin{comment}
We opt not to show the evolution of the GSMF for the HMT and ALS seeding schemes to better underline the differences among $d_{\rm iso}$ cases. In fact, these two models predict stellar mass occupation fraction equal to one already at masses below the exponential cutoff, reproducing the observed GSMF thanks to the AGN feedback as described above. This should not surprise since they are specially implemented to reproduce the GSMF at low redshifts.
\end{comment}

\subsection{Local black hole mass function}
\label{sec:bhmf}
\begin{figure*}
	\includegraphics[width=0.7\textwidth]{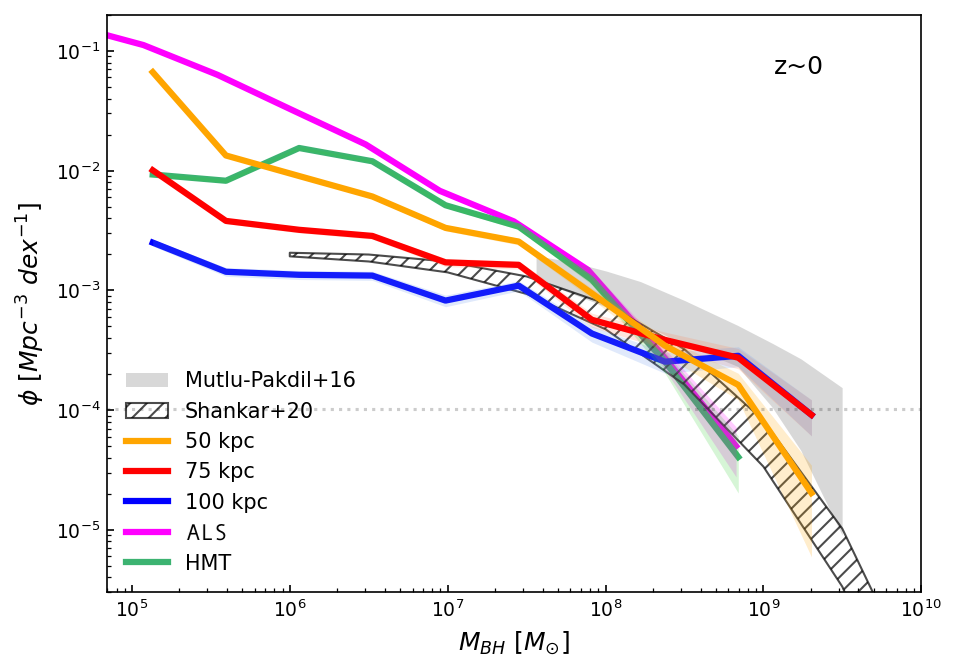}
    \caption{Black Hole Mass Function at redshift $z\sim0$. Solid lines indicate the contribution of different seeding schemes. The shaded areas around the lines depict the Poisson uncertainties in each mass bin. Observed data are taken from a sample of local galaxies \citep[shaded area ][]{Mutlu-Pakdil16}) and from the unbiased corrected relation from \citet{Shankar20} (hatched area). As a reference, the dotted horizontal line stands for 10 objects per mass bin in the whole volume of the box. }
    \label{fig:bhmf}
\end{figure*}

Here we present basic predictions for the mass distribution of the SMBH population at $z \sim 0$. As shown in Fig.~\ref{fig:bhmf}, the three $d_{\rm iso}$ cases predict different trends for the Black Hole Mass Function (BHMF). As a reference, observational constraints are compared to our predictions, derived from a sample of local galaxies \citep[shaded area ][]{Mutlu-Pakdil16} and from the unbiased corrected relation from \citet{Shankar20} (hatched area). Overall, we see that the $d_{\rm iso}=75\:$kpc case predicts number densities of SMBHs in approximate agreement with the observational data. The $d_{\rm iso}=100\:$kpc case falls slightly below the observed values, while the $d_{\rm iso}=50\:$kpc case is slightly higher.

%a lower space density of low-mass SMBH, with respect to the observed relations, while both 50 and 75 kpc models predict a larger space density which tend to align with the extrapolated trend from the observations. 
%As previously discussed, the higher the $d_{\rm iso}$, the more likely the seeds are to merge with gas rich (non-seeded) galaxies. 
%This also means accreting a higher content of gas in the BH reservoir, then available for accretion. In the massive end, this explains the larger space densities of $\sim 10^9 M_{\odot}$ SMBHs. 

%Overall, we find that a reasonable agreement, within uncertainties, between the predictions of our Pop III.1 models and the available constraints would be reached with an  $d_{\rm iso}$ value ranging from 50 to 75 kpc. 

In the high mass end, we see that both the HMT and ALS models do not produce as many very high mass SMBHs as the Pop III.1 models. We expect this is due to a combination of factors: 1) competition between SMBHs for available gas mass; 2) a greater degree of AGN radio-mode feedback, due to larger occupation fractions, which reduces the amount of gas available for SMBH accretion.
%many massive BHs. Note that the HMT and ALS schemes struggle to reproduce the massive end as their host galaxies run out of gas available for accretion due to the higher occupation fractions meaning that more galaxies are subject to the AGN radio-mode feedback.
Conversely, the HMT and ALS models predict the presence of much larger populations of lower-mass SMBHs. In particular, the ALS model, which invokes seeds down to stellar mass scales, predicts a large population of IMBHs.

%Next generation gravitational wave experiments like LISA will help to confirm or discard this possibility.  

%Given the large uncertainties in the observations and the sparse sampling effects affecting both the slope at low BH masses ($\lesssim 10^{7} M_{\odot}$) and the massive end of the BHMF, more robust detection of SMBHs across the mass spectrum could significantly aid in distinguishing between different seeding schemes.

\subsection{The \texorpdfstring{$M_{\rm BH}-M_{\rm star}$}{} relation}
\label{sec:bh_mstar}

\begin{figure*}
    \includegraphics[width=\textwidth]{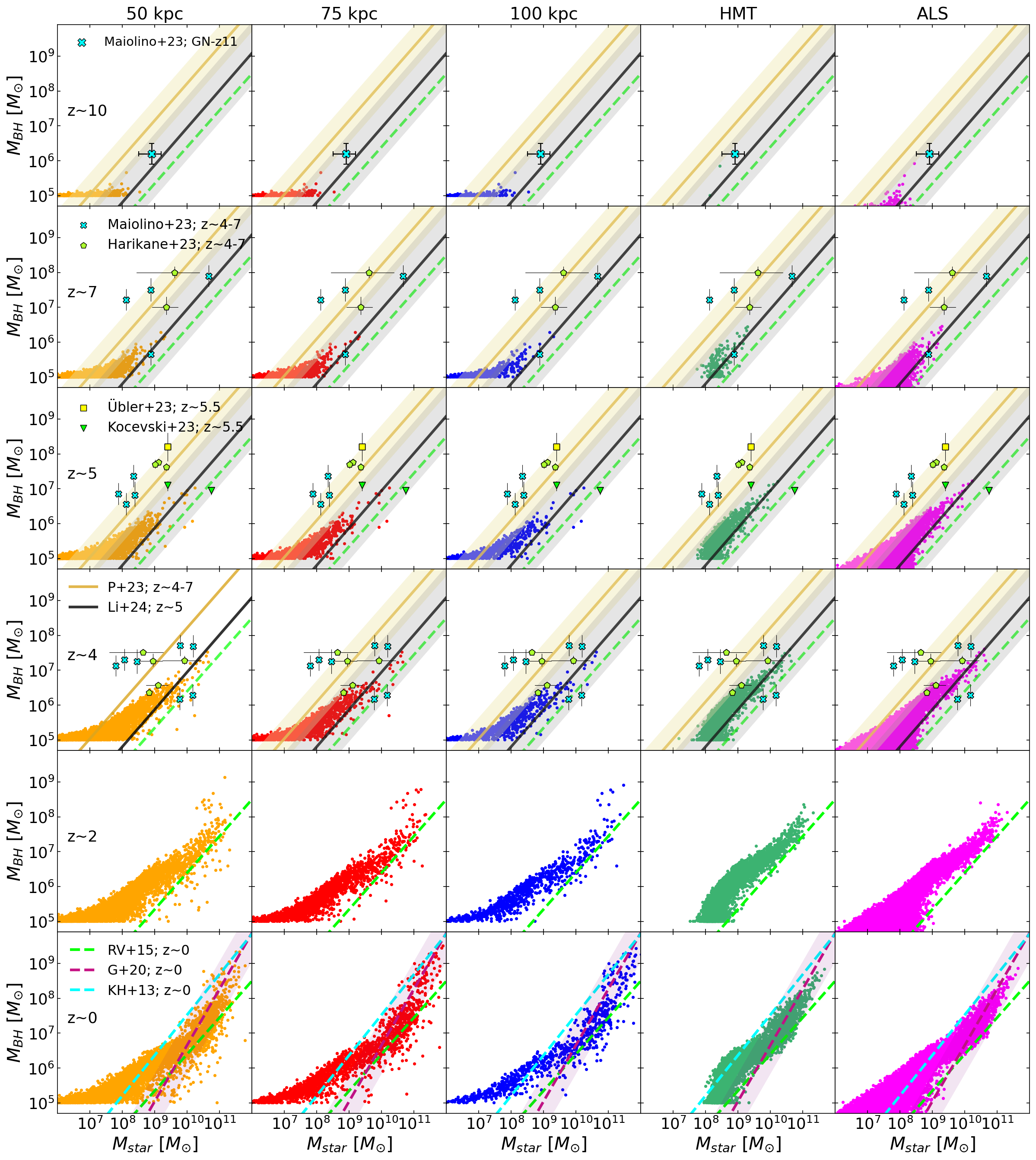}
    \caption{Scatter plot of the evolution of the $M_{\rm BH}-M_{\rm star}$ relation at several redshifts. Different columns represent various seeding schemes. At redshift $z\sim0$ we compare our predictions for our seeded galaxies with several best fits from local observations: dynamical measurements from \citet{Kormendy13} (KH+13), including massive bulge-dominated quiescent galaxies, while \citet{ReinesVolonteri15} (RV+15) use a different mass-to-light ratio, plus an extended sample combining local early- and late-type galaxies by \citet{Greene20} (G+20). At $z\sim2$, we report the RV+15 fit for reference. Moving upward, the high-z rows show the comparison with recent results from the JFAINT sample: in \citet{Pacucci23}, they directly fit the data in the redshift range $z\sim4-7$ while \citet{Li24} estimate an unbiased fit taking into account the uncertainties on the mass measurements and selection effects. Shaded regions illustrate the intrinsic scatter at 1-sigma according to each relation. Colored symbols show faint AGNs taken from \citet{Maiolino23b, Harikane23, Ubler23, Kocevski23} and reported according to their redshifts. In the top row, results at $z\sim10$ are shown against the single data point (GN-z11) from \citet{Maiolino23a}.}
    \label{fig:bh_mstar}
\end{figure*}

\begin{figure}
\centering
	\includegraphics[width=\columnwidth]{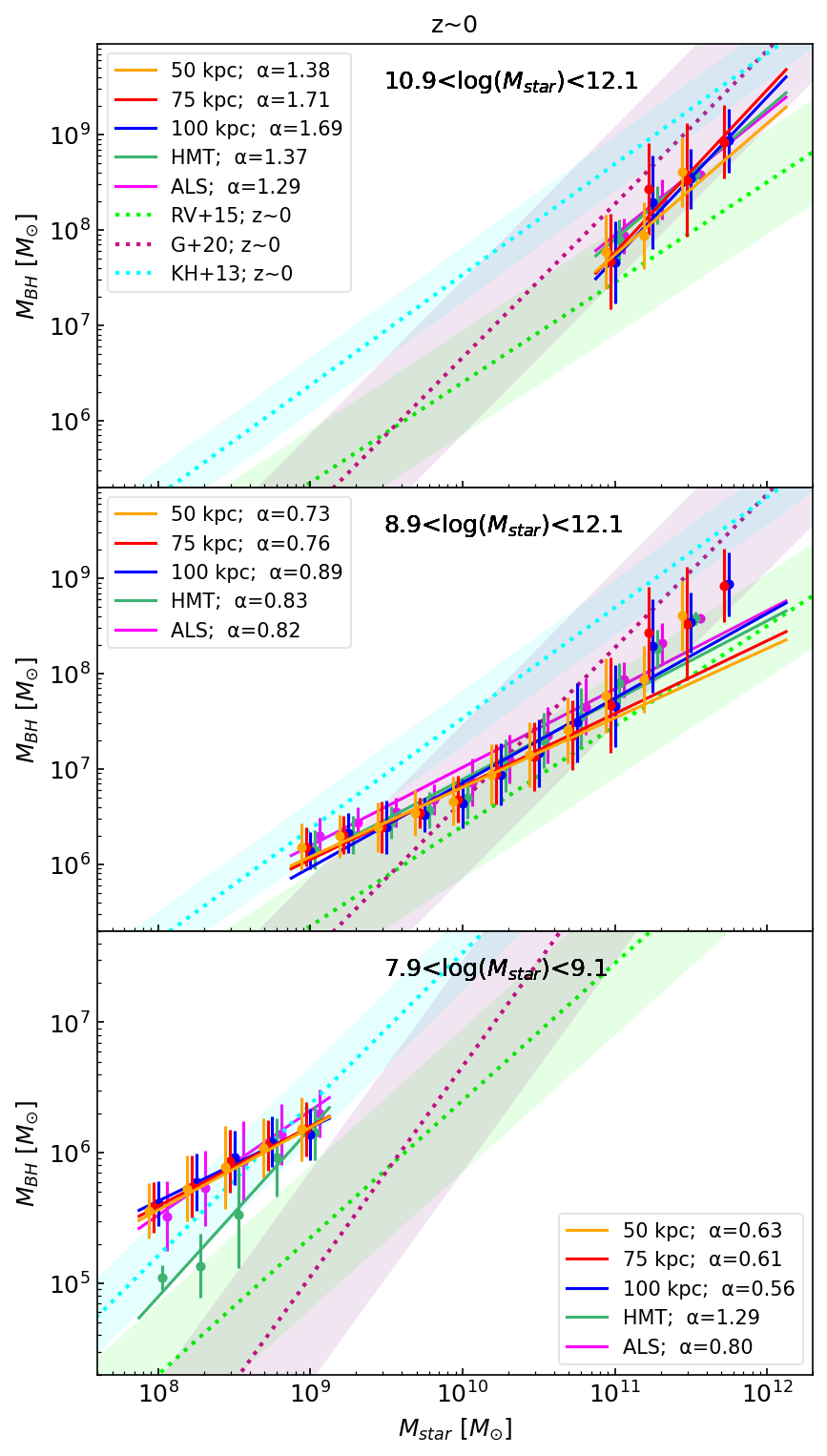}
    \caption{Power law fits to the $M_{\rm BH}-M_{\rm star}$ relation at $z\sim0$. Dotted lines depict local fits with associated dispersion (shaded areas). The indices $\alpha$ are shown in the legend. Solid lines indicate the fit results to the median SMBH masses (data points with dispersion) in several stellar mass bins for different seeding models. The stellar mass range covered by the solid lines corresponds to the fitted interval.}
    \label{fig:fit_bh_star}
\end{figure}

\begin{figure}
\centering
	\includegraphics[width=\columnwidth]{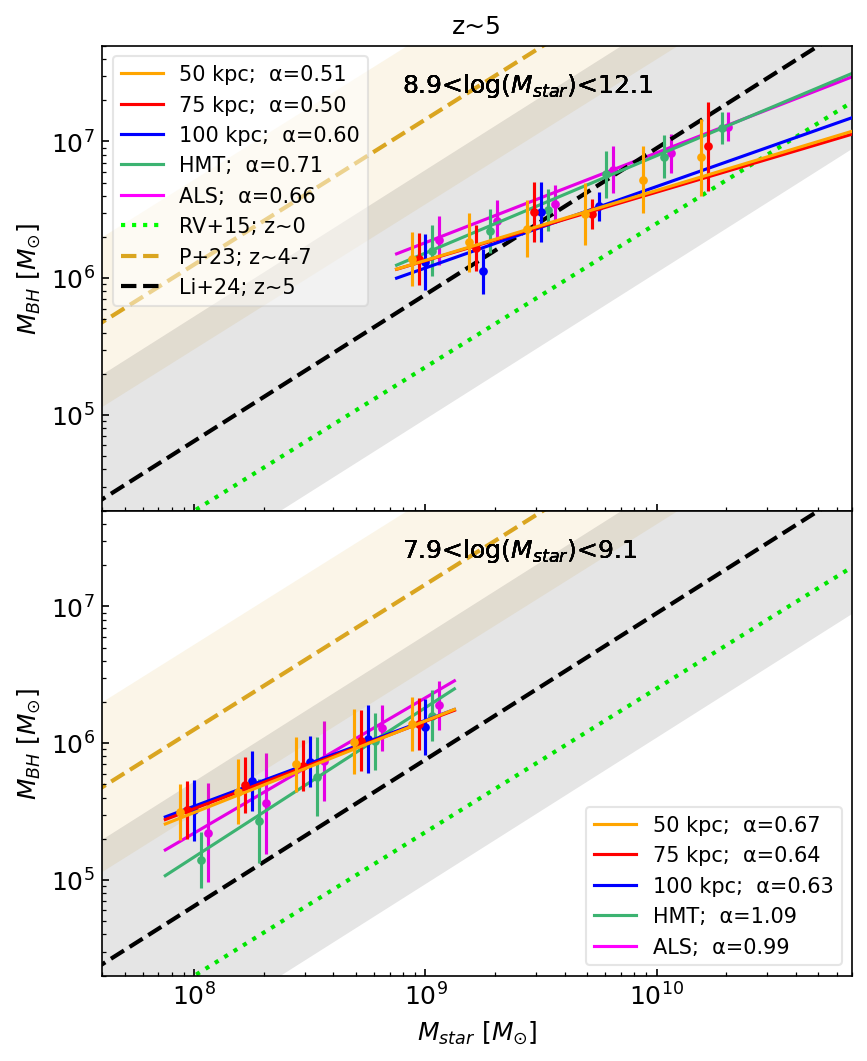}
    \caption{Power law fits to the $M_{\rm BH}-M_{\rm star}$ relation at $z\sim5$. Dashed lines depict fits to the JFAINT data with associated dispersion (shaded areas). For reference, we also show the dotted line reporting the RV+15 fit. The indices $\alpha$ are shown in the legend. \textit{(a) Upper panel:} Solid lines indicate the fit results to the median SMBH masses (data points with dispersion) in several stellar mass bins for different seeding models. The stellar mass range covered by the lines corresponds to the fitted interval in the global range. \textit{(b) Lower panel:} As (a), but the fits extend over the low-mass range.}
    \label{fig:fit_bh_star_z5}
\end{figure}

\begin{table}
    \centering
    \caption{Fitting parameters and derived quantities: the slope $\alpha$, the intercept $M_{\rm BH,9}$ (at $M_{\rm star}=10^{9}M_\odot$), and dispersion $\sigma$ are shown at z$\sim$0. We also report the average (geometric mean) mass of the 10 most massive BHs.}\label{tab:bh_star_z0}
    \begin{tabular}{lccccc}
        & & \textbf{z$\sim$0} & & \\
        \toprule
        & \textbf{50 kpc} & \textbf{75 kpc} & \textbf{100 kpc} & \textbf{HMT} & \textbf{ALS} \\
        \midrule
        $\alpha_{\rm vh}$ & 1.38 & 1.71 & 1.69 & 1.37 & 1.29 \\
        log($M_{\rm BH,9,vh}[M_{\odot}]$) & 4.99 & 4.35 & 4.31 & 5.16 & 5.37 \\
        $\sigma_{\rm vh}[\rm dex]$ & 0.74 & 0.98 & 0.79 & 0.32 & 0.29 \\
        \midrule
        $\alpha_{\rm high}$ & 0.73 & 0.76 & 0.89 & 0.83 & 0.82 \\
        log($M_{\rm BH,9,high}[M_{\odot}]$) & 6.08 & 6.05 & 5.97 & 6.06 & 6.19 \\
        $\sigma_{\rm high}[\rm dex]$ & 0.61 & 0.68 & 0.63 & 0.43 & 0.40 \\
        \midrule
        $\alpha_{\rm low}$ & 0.63 & 0.61 & 0.56 & 1.29 & 0.80 \\
        log($M_{\rm BH,9,low}[M_{\odot}]$) & 6.19 & 6.20 & 6.19 & 6.18 & 6.32 \\
        $\sigma_{\rm low}[\rm dex]$ & 0.50 & 0.43 & 0.40 & 0.50 & 0.51 \\
        \midrule
        log($\overline{M}_{\rm BH,max10}[M_{\odot}]$) & 8.99 & 9.26 & 9.23 & 8.62 & 8.65 \\
        \bottomrule
    \end{tabular}
\end{table}

\begin{table}
    \centering
    \caption{Fitting parameters and derived quantities: the slope $\alpha$, the intercept $M_{\rm BH,9}$ (at $M_{\rm star}=10^{9}M_\odot$), and dispersion $\sigma$ are shown at z$\sim$5. We also report the average (geometric mean) mass of the 10 most massive BHs.}\label{tab:bh_star_z5}
    \begin{tabular}{lccccc}
        & & \textbf{z$\sim$5} & & \\
        \toprule
        & \textbf{50 kpc} & \textbf{75 kpc} & \textbf{100 kpc} & \textbf{HMT} & \textbf{ALS}\\
        \midrule
        $\alpha_{\rm high}$ & 0.75 & - & - & 0.83 & 0.78 \\
        log($M_{\rm BH,9,high}[M_{\odot}]$) & 6.08 & - & - & 6.13 & 6.19 \\
        $\sigma_{\rm high}[\rm dex]$ & 0.46 & - & - & 0.39 & 0.36 \\
        \midrule
        $\alpha_{\rm low}$ & 0.65 & 0.60 & 0.58 & 0.96 & 1.04 \\
        log($M_{\rm BH,9,low}[M_{\odot}]$) & 6.07 & 6.02 & 6.01 & 6.13 & 6.23 \\
        $\sigma_{\rm low}[\rm dex]$ & 0.39 & 0.38 & 0.35 & 0.49 & 0.58 \\
        \midrule
        log($\overline{M}_{\rm BH,max10}[M_{\odot}]$) & 6.75 & 6.65 & 6.56 & 6.91 & 6.93 \\
        \bottomrule
    \end{tabular}
\end{table}

Here we consider the relation between central SMBH mass and galactic stellar mass for the different models of SMBH seeding. Recall that, as described above, these results are based on the fiducial SMBH growth model adopted by F20. The scatter plots in Fig.~\ref{fig:bh_mstar} show the populations of seeded galaxies for the various seeding schemes (columns) at redshift zero (bottom row), with high redshift results shown in the other rows. Note that in the following analysis, we exclude unseeded galaxies, which, as discussed in \S\ref{sec:occ_fract}, for the Pop III.1 models can be a significant fraction of the systems even in the high $M_{\rm star}$ regime. Thus when comparing our results to observational samples, only galaxies with confirmed SMBHs, i.e., not just upper limits, should be used.

\subsubsection{Redshift zero results}

Focusing first on the local results, we see a general trend of more massive black holes being found in galaxies with larger stellar masses. Since the Pop III.1 model assumes seed masses of $10^5\:M_\odot$, we see a flattening of the $M_{\rm BH}-M_{\rm star}$ relation to this level as one goes to lower mass galaxies. This feature is not seen in the HMT model since requiring a threshold halo mass of $7\times 10^{10}\:M_\odot$ naturally imposes an effective threshold on $M_{\rm star}$ for seeded galaxies. This feature is also not seen in the ALS seeding scheme, since the seed masses in this model are much smaller than $10^5 M_{\odot}$. 

In general, the $M_{\rm BH}-M_{\rm star}$ relation can be approximated as a power law, especially if one restricts to limited ranges in stellar mass. We thus fit the following function to the distributions:
\begin{equation}
    M_{\rm BH} = M_{\rm BH,9} \left(\frac{M_{\rm star}}{10^9\:M_\odot}\right)^{\alpha}
\end{equation}
and consider a stellar mass of $10^9\:M_\odot$ as fiducial scale at which to divide \textit{low-mass} and \textit{high-mass} galactic systems. However, we set a lower limit of $M_{\rm star}=10^8\:M_\odot$, which is designed to make the metrics of the low-mass case easier to compare to observed systems. We also consider a \textit{very high-mass} range 
%labelled as \textit{vh}, 
at $M_{\rm star}>10^{11}\:M_\odot$. We carry out power law fits to the binned median values of $M_{\rm BH}$, i.e., giving equal weight to the different bins of $M_{\rm star}$, which are distributed evenly in logarithm and spaced every $\sim$0.33 dex. We require at least 5 sources in a given bin to include it in the fit. Every median value is then weighted with its corresponding standard error, which in the least populated high-mass bins is about 0.3 dex. We report the maximum range of the black hole distribution via a metric $\overline{M}_{\rm BH,max10}$, which is the geometric mean mass of the ten most massive black holes in the simulation domain of $\sim (60\:{\rm Mpc})^3$. We also measure the dispersion, $\sigma$ about each best-fit power law $M_{\rm BH}-M_{\rm star}$ relation, averaging the dispersions in each mass bin equally. The redshift zero results for $\alpha$, $M_{\rm BH,9}$ and $\sigma$ for the low-, high- and very high-mass cases, as well as $\overline{M}_{\rm BH,max10}$, are reported in Table~\ref{tab:bh_star_z0} for the various seeding schemes. These power law fits are also plotted in Fig.~\ref{fig:fit_bh_star}.

%We see that 
%Pop III.1 models illustrate similar trends as a function of isolation distance $d_{\rm iso}$ in terms of both the distribution and the scatter of the data. In particular, 
%there are two main differences among our z$\sim$0 predictions. First, 
%the properties of galaxies hosting black holes of $M_{\rm BH} \sim 10^{5-6} M_{\odot}$ depend strongly on $d_{\rm iso}$. For smaller $d_{\rm iso}$ such black holes can be found in galaxies with much larger stellar masses. 
%Or, considering $M_{\rm star}$ as the dependent variable, 

For galaxies with $M_{\rm star}<10^9\:M_\odot$, the properties of the best fit power law show that the Pop III.1 models have the most shallow power law indices, i.e., $\alpha_{\rm low}\simeq 0.6$, with the $d_{\rm iso}=100\:$kpc case having the shallowest index of 0.56. In comparison, the HMT model has $\alpha_{\rm low}\simeq 1.3$, while the ALS model has $\alpha_{\rm low}\simeq 0.8$. On the other hand the amplitude of the power law fits, as measured by $M_{\rm BH,9}$, show very little variation ($\lesssim0.1$ dex) between the models. The dispersion about the power law is similar among the models at 0.5~dex, except for the two larger $d_{\rm iso}$ cases where it drops at 0.4~dex for the 100~kpc model.
%and the ALS case have shallower slopes, while the HMT model depict a steeper coefficient. This is due the limit in halo mass, i.e. in stellar mass, that the HMT model poses on the galaxies to be seeded. 
This trend in the Pop III.1 models results from the fact that for smaller values of $d_{\rm iso}$, a broader variety of halos is seeded, leading to a wider range of growth and star formation histories that then lead to greater scatter in the $M_{\rm BH}-M_{\rm star}$ relation.
%galaxies are seeded across a broader mass spectrum, which flattens out their relation in this mass regime, while the ALS scheme provides an intermediate slope since there's no plateau for the mass of the seeds. 
%This is explain the larger dispersion observed in smaller $d_{\rm iso}$ cases, comparable with the HMT and ALS schemes. 
%In terms of the amplitude $M_{\rm BH,9}$, all models present similar values around few in $10^6\:M_\odot$. 

%In fact, by $M_{\rm star}\sim10^8\:M_\odot$ 

In the high-mass regime, with $M_{\rm star}>10^9\:M_\odot$, the processes leading to black hole growth and star formation result in an $M_{\rm BH}-M_{\rm star}$ relation that is relatively similar between the different seeding schemes, so it becomes harder to distinguish the models from the statistics of their $M_{\rm BH}-M_{\rm star}$ relations. Nevertheless, from the results shown in Table~\ref{tab:bh_star_z0} we notice that, in contrast to the low-mass regime, the Pop III.1 models have systematically higher values of dispersion about their best-fit power laws, i.e., $\sigma_{\rm high}\simeq 0.7$, while the HMT model has $\sigma_{\rm high}\simeq 0.5$ and the ALS model 0.4. Similar conclusions can be drawn for the very high-mass regime. This is caused by the Pop III.1 models having quite large volumes that are prevented from forming seeds, i.e., via the isolation distance criteria, and these volumes can contain relatively massive halos that form galaxies with relatively high $M_{\rm star}$ that either never host a SMBH (see the comparatively low occupation fractions in Fig.~\ref{fig:occ_fract_zoom} of the Pop III.1 models) or gain a relatively low-mass SMBH from later merger with a smaller halo/galaxy. %\textbf{\textcolor{red}{We also note that our reported values of scatter in the $M_{\rm BH}-M_{\rm star}$ relation for the Pop III.1 models ignore the impact of unseeded galaxies, and in this sense are underestimates. As discussed in \S\ref{sec:occ_fract}, there are significant populations of unseeded galaxies with relatively high values of $M_{\rm star}$ in the Pop III.1 models, especially with larger values of $d_{\rm iso}$.}}

We also notice that at the highest masses there is evidence for steepening of the power law behavior of the $M_{\rm BH}-M_{\rm star}$ relation. This is most noticeable for the Pop III.1 models, which tend to have a greater number of more massive SMBHs (see Fig.~\ref{fig:bhmf}). After inspecting the accretion histories of some example cases, we attribute this as being due to reduced impact of AGN feedback in the Pop III.1 models given their overall smaller numbers of SMBHs. Indeed, AGN feedback processes act as regulators of the cold gas content in galaxies, by heating their cold gas component and displacing it to the hot phase. This implies that unseeded galaxies in Pop III.1 models tend to accumulate larger cold gas reservoirs with respect to seeded ones. Therefore, when these galaxies became satellites in massive haloes and merge with a central seeded galaxies, they provide more cold gas available for accretion onto the central SMBH with respect to seeded galaxies (see \S\ref{subsec:accretion}), thus enhancing the final BH mass achieved compared to the HMT and ALS cases.

To quantify these differences, we refer to the power law fits for the very high-mass regime (see Table~\ref{tab:bh_star_z0}).
In the top panel of Fig.~\ref{fig:fit_bh_star} we see that for larger values of $d_{\rm iso}$ Pop III.1 models show a steeper slope up to $\sim$1.7 for the 100 and 75 kpc cases, while other seeding criteria present comparable shallower trends with power law indexes about 1.3-1.4. This is a direct consequence of the already mentioned competition effect on the BH accretion (e.g., see also the BHMF in Fig.~\ref{fig:bhmf}). In terms of scatter, we observe larger values in this stellar mass regime for the Pop III.1 model. 

%jct - I don't think we should say this:
%In fact, given the poorer statistics, we expect a non negligible contribution to the measured scatter from outliers, which are more frequent in Pop III.1 cases. 

%the accretion onto the SMBH homogenises their mass, losing memory of the initial value of the seed mass. Similar conclusions can be drawn for the amplitudes of the high-mass fits, as well as for the slopes of the different seeding models where we do not predict significant differences. However, we see that the dispersion shows notable variation in the Pop III.1 cases with respect to the HMT and ALS models. 

Related to the very-high mass regime of the $M_{\rm BH}-M_{\rm star}$ relation, we see that $\overline{M}_{\rm BH,max10}$ is higher in the Pop III.1 models, i.e., $1-2\times 10^9\:M_\odot$, than in the HMT or ALS models, where it is $\lesssim 5\times 10^8\:M_\odot$. We again attribute this to reduced competition for gas in the Pop III.1 models due to smaller overall numbers of SMBHs. Also, the increased number of BH mergers in HMT and especially in ALS is not significant for the overall growth of the central SMBH, as the majority of mergers can contribute only up to a few percent of the total mass. Recall that the extra seeds in this two schemes are likely to happen in relatively low-mass halos since the most massive ones are seeded in all the scenarios.

We compare our $z=0$ results to various observational constraints (see Fig.~\ref{fig:fit_bh_star}). In particular, we consider the empirical $M_{\rm BH}-M_{\rm star}$ relations obtained from: 1) the $M_{\rm BH}-M_{\rm bulge}$ relation from \citet{Kormendy13} (hereafter KH+13), who use dynamical measurements of massive bulge-dominated quiescent galaxies and who find an intrinsic scatter about the relation of about 0.3 dex; 2) the \citet{ReinesVolonteri15} (hereafter RV+15) relation, which fits the same $M_{\rm BH}-M_{\rm star}$ data set of KH+13, but with an updated mass-to-light ratio, resulting in a lower normalization; 3) the \citet{Greene20} (hereafter G+20) relation, which considers low-mass ($\sim10^{5} M_{\odot}$) SMBHs and their host galaxies.
We note that the inclusion of low-mass objects in the G+20 fit down to stellar masses of $\sim10^9 M_{\odot}$ causes the sample to be sparse and potentially biased for stellar masses below $10^{10} M_{\odot}$, where the number of objects measured via dynamical methods is small. We note also that low-mass and faint SMBHs are more likely to be missed in observational surveys, potentially biasing the derived relations \citep[e.g.,][]{Shankar20}. 

In the range of measured stellar masses from $\sim10^{9} M_{\odot}$ up to $10^{12} M_{\odot}$, all theoretical predictions show agreement within the 1-sigma bands of the observational fits provided by RV+15 and by the sample used in G+20. As discussed above, the largest differences between the theoretical models in this regime are in the very high-mass regime, so this may be a promising area for future, more detailed observational tests.

In the low-mass regime, i.e., $\lesssim10^{9} M_{\odot}$, there are larger differences between the models. As discussed the Pop III.1 models have shallower indices and higher amplitudes than the HMT and ALS models, and these appear to be less in agreement with the RV+15 and G+20 observational results. However, as mentioned the data are relatively sparse in this regime and potentially subject significant systematic uncertainties and observational biases. From the theoretical point of view the precise amplitude and power law index is also sensitive to choice of a single seed mass of $10^5\:M_\odot$. Still, since the models show greatest differences in this regime, we consider than improving the observational constraints at these lower values of $M_{\rm star}$ is a high priority.

\subsubsection{High-redshift results}

Moving to higher redshifts, in the first five rows of Fig.~\ref{fig:bh_mstar} we show the evolution of the $M_{\rm BH}-M_{\rm star}$ relation of the simulated seeded galaxies from $z\sim10$ down to $z\sim2$.
%, with particular focus in the redshift interval from 4 to 7. 
To guide the eye, the light green dashed line shows the local relation from RV+15. Table~\ref{tab:bh_star_z5} reports power law fit results for the $z\sim 5$ sample.

As we proceed to higher redshift, the most obvious feature is that there are relatively few massive galaxies and associated highest-mass SMBHs. To help quantify this trend, Table~\ref{tab:bh_star_z5} reports the results for ${\rm log} \overline{M}_{\rm BH,max10}$ for the different seeding schemes at $z\sim5$. We see that ${\rm log} (\bar{M}_{\rm BH,max10} \simeq 7$ for most of the models, dropping to 6.56 for the Pop III.1 model with $d_{\rm iso}=100\:$kpc. This is likely to reflect the fact that this case has the fewest SMBHs and so a reduced sampling of the relatively rare conditions that lead to the strongest growth. In the high-mass regime, the $d_{\rm iso}=75$ and 100~kpc cases do not form sufficient SMBHs for us to measure the $M_{\rm BH}-M_{\rm star}$ relation. For the other seeding schemes, we see quite similar power law fits, but with the Pop III.1 $d_{\rm iso}=50\:$kpc case having moderately higher dispersion. In the low-mass regime, as at $z=0$, we see significantly shallower indices in the Pop III.1 models ($\alpha_{\rm low}\simeq 0.6$) compared to the HMT and ALS models ($\alpha_{\rm low}\simeq 1.0$. Furthermore, the Pop III.1 models have smaller dispersions than the HMT and ALS cases. Among the different Pop III.1 models, the effect of varying the isolation distance $d_{\rm iso}$ is relatively hard to distinguish from the $M_{\rm BH}-M_{\rm star}$ relation fits.

%is even less noticeable than at low $z$, since most of the accreting SMBHs reside in the most massive halos, which in turn are seeded regardless of the isolation criterion in the considered range of $d_{\rm iso}$. 

Considering the evolution with redshift, among the various seeding schemes, the HMT is the one showing the largest evolution from $z\sim10$ down to 4 as the process of seeding halos with SMBHs starts relatively late (i.e., $z\sim10$, see Paper II). In the ALS seeding model, the BH seeds grow relatively fast and by $z\gtrsim7$ they have already caught up with the SMBH populations formed from heavy seeds. This is due to the combination of the viscous accretion mode onto the BH and the small seed mass. In fact, seeding with stellar mass BHs allows these objects to accrete several times their own mass as the BH accretion rate goes up to 10 times the Eddington limit and they keep accumulating mass until they become massive enough to self regulate their own growth primarily via AGN radio-mode feedback (see also Fig.~\ref{fig:edd_ratios_z}). 

In Fig.~\ref{fig:bh_mstar} we also compare the theoretical models with recent results obtained from deep JWST galaxy surveys where it has been possible to find very faint AGNs otherwise undetected with other facilities (hereafter JFAINT sample). Due to the relatively small volume of our \pinocchio box ($\sim 60:{\rm cMpc}$ per side; see Section~\ref{subsec:pinocchio}), caution is required when comparing these theoretical results with the full JFAINT sample, which is derived from surveys covering larger volumes. In particular, larger simulation volumes  are required to make comparison with the rarest, very luminous quasars. 

Focusing on redshifts 4 to 7, we plot data from \citet{Maiolino23b} (twelve low-luminosity AGNs), \citet{Harikane23} (ten objects), \citet{Kocevski23} (2 objects) and \citet{Ubler23} (1 object). This JFAINT sample includes sources available from different surveys, specifically from the CEERS \citep{Finkelstein23}, JADES \citep{Eisenstein23} and ERO \citep{Pontoppidan22} programs. In the $z\sim$10 row we report the single data point (GN-z11) from \cite{Maiolino23a}. It is worth noting that the JFAINT sample has been selected by looking for the broad component of the H$\alpha$ and H$\beta$ lines by means of NIRspec requiring a specific threshold for the full width half maximum (FWHM). Additionally, BH masses are measured using a relation connecting the BH mass to the FWHM of the broad line, mainly H$\alpha$, calibrated locally using the decomposition of the broad and narrow components of the emission lines \citep[see ][]{Maiolino23a, Harikane23}. Stellar masses are instead typically estimated via SED fitting, which carries significant uncertainties to corrections related to dust attenuation, assumptions about metallicity, and AGN contribution.

This sample of JWST-detected AGNs has been used to estimate the $M_{\rm BH}-M_{\rm star}$ at earlier epochs. \citet{Pacucci23} (P+23) performed a direct fit of the JFAINT data in order to derive the intrinsic $M_{\rm BH}-M_{\rm star}$ relation at a mean redshift of $\sim 5$ (golden solid line), concluding that high-$z$ SMBHs tend to be overmassive by a factor of 10-100 with respect to the local relation. This interpretation implies that, even with very massive seeds, super-Eddington accretion episodes are required to frequently occur at high redshift. More recently, \citet{Li24} (Li+24) have presented a study of the same sample of objects as in P+23 including a detailed analysis of the possible biases. In fact, uncertainties due the measurement of both BH and stellar mass and selection effects caused by flux-limited detection may lead to biased conclusions. According to these authors, the observed data can be explained by assuming that the intrinsic $M_{\rm BH}-M_{\rm star}$ relation is more similar to the local one (e.g., of KH+13). Their result, shown by the black line, suggests that local and high-z relations behave in a similar way, where the JFAINT sample is an extremely biased selection towards the most luminous objects. 

%This can explain the difference with the analysis of P+23. 
%Bearing in mind all these caveats is crucial to study and contextualize the evolution of the  $M_{\rm BH}-M_{\rm star}$ relation throughout the cosmic history when comparing simulations and observations.

Our predicted BH masses are, on average, lower by one order of magnitude in BH mass with respect to the most massive BHs in the JFAINT sample (with mean redshift $\simeq5.2$), while we can reach reasonably high stellar masses by redshift $\sim5$. Our $M_{\rm BH}-M_{\rm star}$ relations do not match the P+23 fit, while showing a much better agreement with the estimate of Li+24. If the P+23 results are confirmed as a robust estimate for the true high-$z$ $M_{\rm BH}-M_{\rm star}$ relation, there are two possible arguments that can explain the lack of very massive BHs in our model predictions. On the one hand the accretion scheme onto the central BH used in this study was calibrated by F20 in the \gaea framework to match AGN bolometric luminosity functions up to redshift $\sim4$. The fact that our realizations do not recover such high BH masses may be connected with the physics implemented in the growth model driven by gas momentum loss and viscous accretion, which sets the timescale of the accretion rate being inversely proportional to the BH mass as in Eq.~\eqref{eq:rate_bh}. However, at higher redshift cold gas accretion may be expected to be a more continuous process \citep[e.g., ][]{Inayoshi16} that does not need trigger events such as galaxy mergers and/or disc instabilities to happen. If true, this could lead SMBHs to grow on shorter timescales and/or at higher efficiencies than we have so far implemented (see the discussion in \S\ref{subsec:accretion}). 

Additionally, the limited size of our \pinocchio box also limits our ability to sample extreme, rare objects that undergo the fastest, most efficient accretion.
%build up in terms of BH mass as well as in the scatter of the data points. 
This volume effect is particularly evident by looking at the top row at $z\sim10$, where only single objects can occasionally match the GN-z11 data point within its error bars.
%(see top row in Figure~\ref{fig:bh_mstar}), but larger statistics are needed to draw more robust conclusions. 
We note that our simulated volume is at least a factor of 10 smaller than those probed by the JADES and CEERS programs.

Furthermore, the measurements of BH masses using local calibrations may be significantly biased upward (by up to one order of magnitude), as suggested recently by \citet{Abuter24}. Other considerations are that for the Pop III.1 and HMT models we seed galaxies with a delta function distribution of BH mass at $10^5 M_{\odot}$. More realistically, we would include a range of masses with dispersion, and this would enhance the chance to obtain a few, larger SMBH masses. 

%We plan to further investigate the comparison with recent data from JWST as well as different implementations of BH accretion mode in a future work. 

%Finally, we attempt to fit our simulated data in the intermediate high redshift z$\sim$5 across the high-mass range, finding that very mild differences arise in terms of the fit coefficients. As for z$\sim$0, the HMT depicts a steeper slope due to the threshold mechanism which does not seed low stellar mass galaxies. Again, with current observational data, it will be arduous to discriminate among different seeding criteria.

\subsection{Eddington ratios}\label{subsec:edd_ratios}

\begin{figure}
    \includegraphics[width=\columnwidth]{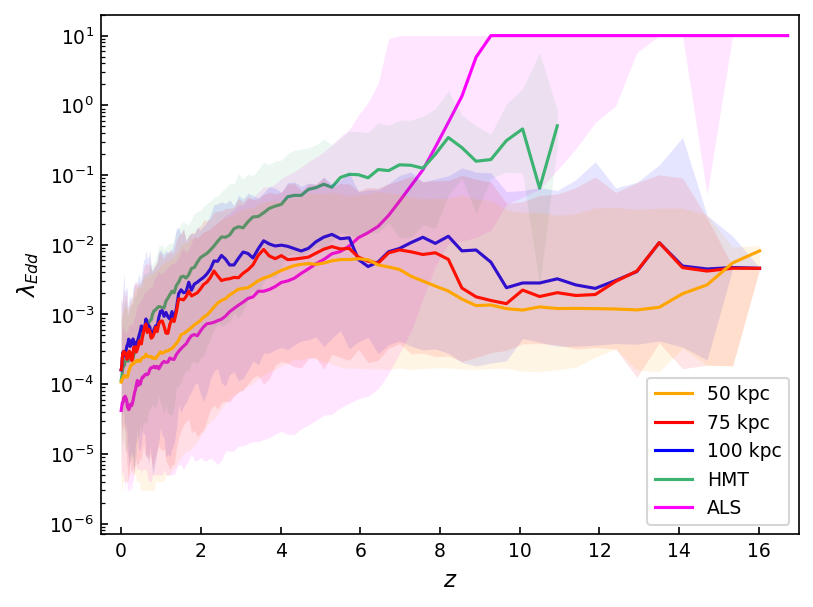}
    \caption{Medians (in log scale, solid lines) of the Eddington ratio $\lambda_{\rm edd}$ as a function of redshift for the different seeding mechanisms. Shaded areas denote the 1-sigma dispersion of the distributions.}
    \label{fig:edd_ratios_z}
\end{figure}

\begin{figure*}
    \includegraphics[width=\textwidth]{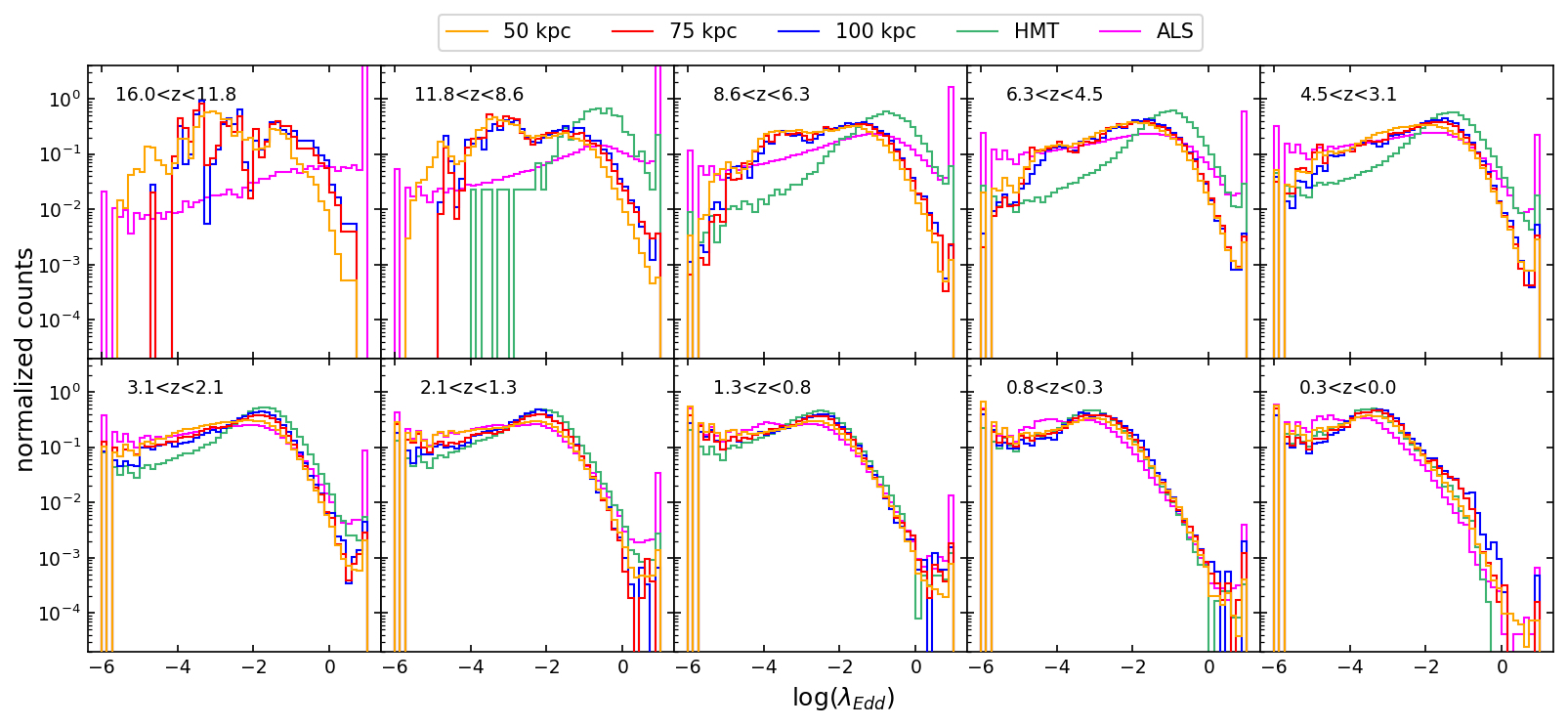}
    \caption{Normalized histograms (in log scale) of the Eddington ratio $\lambda_{\rm edd}$ for in different redshift bins as a function of the seeding mechanism. The spacing in redshift bins corresponds to a linear spacing in the logarithm of the scale factor $a$. }
    \label{fig:edd_ratios}
\end{figure*}

%We also attempt to give a quantitative explanation for the trends exhibited by the $M_{\rm BH}-M_{\rm star}$ relations by looking at the Eddington ratios $\lambda_{\rm edd}$. 

Here we examine the distribution of Eddington ratios, $\lambda_{\rm edd}$, that are present in our modeled SMBH populations. Figs.~\ref{fig:edd_ratios_z} and \ref{fig:edd_ratios} show in different formats the distribution of $\lambda_{\rm edd}$ across different seeding models as a function of redshift. The solid lines in Fig.~\ref{fig:edd_ratios_z} show the median $\lambda_{\rm edd}$ of the different models, with shaded areas being the $1\sigma$ dispersion as a function of redshift. Fig.~\ref{fig:edd_ratios} shows the normalized histograms of $\lambda_{\rm edd}$ in ten redshift bins. For Pop III.1 models, the distributions, including median values, of $\lambda_{\rm edd}$ are relatively similar. We do notice that, especially at higher redshifts, the cases with larger $d_{\rm iso}$ have a higher proportion of higher accretion rate SMBHs.
%accretion rates compared to the Eddington limit present similar behaviour: their averages and distributions are relatively similar as $d_{\rm iso}$ varies with a mild effect on smaller $d_{\rm iso}$ having slightly lower $\lambda_{\rm edd}$, as anticipated from the BHMF and the $M_{\rm BH}-M_{\rm star}$ relation. 
Comparing high and low redshifts, at higher redshifts the SMBHs tend to have higher accretion rates, with a median $\lambda_{\rm edd}$ of $\sim 10^{-3}$ to $10^{-2}$ down to $z\sim 1$ for $d_{\rm iso}=75$ and 100~kpc cases and $z\sim 2$ for $d_{\rm iso}=50$~kpc. After this, SMBHs start to self-regulate their growth via AGN feedback. This decreases the available gas which in turn lowers the average accretion to $\lambda_{\rm edd}\sim 10^{-4}$ by $z\sim$0. However, the histograms in Fig.~\ref{fig:edd_ratios} show that the SMBHs have a tail towards accretions close to the Eddington limit (or even super-Eddington), which would manifest as luminous AGN, including quasars.
%, as observed in rare and powerful quasars at intermediate redshifts.

On the other hand, the HMT and ALS models exhibit different distributions of $\lambda_{\rm edd}$. In the case of HMT, the BH seeds appear at lower redshift ($z\sim10$) in already massive halos where a substantial amount of gas available for accretion has gathered during its previous history. At this point, the central BHs are able to accrete efficiently resulting in $\lambda_{\rm edd}$ of about an order of magnitude larger than in the Pop III.1 models. %After a transient period of time, 
However, below $z\sim3$ the HMT BHs tend to align with the behaviour of the Pop III.1 populations. 

For the ALS model, the BH seeds are significantly lower in mass than the other seeding models considered in this study. This explains the fact that down to $z\sim8$ the large majority of BHs are accreting at super-Eddington rates (limited at 10 times as a basic model assumption). In fact, in the viscous accretion prescription the accretion rate is proportional to the ratio between the BH gas reservoir and the BH mass. Until the ALS BHs reach the supermassive regime, this ratio takes values much larger with respect to the other models, explaining the trends observed in Figs.~\ref{fig:edd_ratios} and \ref{fig:edd_ratios_z}. At lower redshifts, these systems tend to run out of gas available for BH accretion, as all halos are seeded (e.g., see the BHMF in Fig.~\ref{fig:bhmf}). We note that by $z\sim2$ and down to the local Universe, all the models have generally similar distributions, although the ALS model retains an excess of very high, super-Eddington accretors, mostly being low- and intermediate-mass BHs.

%Combining the results of the $M_{\rm BH}-M_{\rm star}$ relation and $\lambda_{\rm edd}$, one can better understand the limitations of our current implementation, especially in the context of reproducing high-z over massive BHs. On the one hand the viscous accretion described in Sec.~\ref{subsec:accretion} is not able to make massive BH seeds grow statistically faster than a fraction of percent of the Eddington limit. On the other hand, light seeds as in the ALS model start with too low BH masses and they end up in the same regime of masses of the others models, even if they accrete at super-Eddington ratios.   

\section{Summary \& Conclusions}
\label{sec:conclusions}

In this work we have introduced a novel semi-analytic approach that accounts for different SMBH seeding scenarios within theoretical models of galaxy formation and evolution. It utilizes merger trees generated by the \pinocchio code for dark matter halos, extends them by incorporating subhalos, and then applies the \gaea semi-analytic model to populate these halos with observable galaxies. This approach allows us to investigate a wide range of galaxy properties by adjusting various parameters governing galaxy formation without expensive N-body and/or hydrodynamical simulations. The evolution of subhalos and their merging with the main halos is implemented via physically motivated models that have been calibrated on simulations.   

We first adapted the structure of \pinocchio merger trees to the Millennium simulation format by adding information about the subhalos. We assume a spatial distribution for subhalos following a Navarro-Frenk-White density profile \citep[NFW, ][]{Navarro97}, statistical prescriptions for the angular momentum \citep{Zentner05, Birrer14} and a subhalo survival time since accretion \citep{Berner22, Boylan08}. This ensures that the \gaea model will be able to run on \pinocchio-generated halo merger trees. We calibrated this method on a Millennium-like \pinocchio box making sure that we fed the semi-analytic model with a consistent halo mass function. By anchoring the calibration to the local observed GSMF, we estimated a total survival time for satellite galaxies, which is tuned to reproduce the exponential cut-off of the high-mass end of the predicted GSMF at $z\sim0$. This approach makes it possible to apply our fully semi-analytic pipeline to a variety of scientific cases beyond the scope of the current paper.

We have primarily focused on the impact of implementing different mechanisms for seeding SMBHs, especially focusing on the Pop III.1 model, which postulates a new mechanism for the formation of all SMBHs. We have investigated three values of the isolation distance that is needed for a given minihalo to be a Pop III.1 source. For comparison, we have examined the Halo Threshold Model (HMT) used by the Illustris-TNG simulations in which every halo exceeding a mass of $7.1 \times 10^{10} M_{\odot}$ is seeded with a BH of mass $1.4 \times 10^{5} M_{\odot}$. As another example case, we also considered predictions for the standard seeding scheme implemented in \gaea based on ALS. Here, the initial mass of the BH seed scales with the initial halo mass, resulting in light seeds of the order of stellar mass BHs.
%consistent with Pop III remnants. 

Within our framework, we have explored the implications of this set of seeding models when applied to cosmological volumes in a galaxy formation and evolution framework down to low redshifts.
Unlike other astrophysical models, the Pop III.1 scenario presents the earliest and least clustered distribution of seeds, affording relatively longer periods of time for black hole growth via accretion, reducing the need for 
%and accounting for high-redshift quasars without requiring s
sustained modes of super-Eddington accretion. 

%With only one free parameter, the isolation distance ($d_{\rm iso}$), the Pop III.1 model is relatively straightforward and conducive to exploration in cosmological volume simulations that resolve minihalos, including galaxy formation models like \gaea.

Our main findings are the following:
\begin{itemize}

\item For the Pop III.1 models, SMBH seeds are predominantly abundant in massive galaxies, with the occupation fraction increasing as $d_{\rm iso}$ decreases. By $z\sim0$, the occupation fraction reaches unity for halo masses above $10^{13} M_{\odot}$ across all models. AGN feedback significantly influences the thermal state of the gas and the SFR, quenching seeded galaxies by $z\sim0$.  The observational measurements of the occupation fraction in the local Universe as a function of stellar mass suggest that $d_{\rm iso} < 75\:$ kpc. In contrast, the HMT and ALS scheme produce too many BHs in systems with $M_{\rm star}\sim 10^{9-10} M_{\odot}$.
%, observations estimate occupation fractions about 50\%, in agreement with the $d_{\rm iso}=$ 50 kpc case predictions.

\item The AGN radio-mode feedback affects the shape of the GSMF by decreasing the SFR in massive galaxies, if hosting a SMBH, and causing the exponential cut-off at the high-mass end. At $z\sim0$, the low-mass end of the GSMFs obtained from Pop III.1 models is generally similar to the predictions of the HMT and ALS models and with the observations, as in this regime the number densities are regulated by SN feedback. Above few $\times 10^{13} M_{\odot}$, smaller isolation distances are favoured to reproduce the quenching of the majority of massive galaxies, consistent with $d_{\rm iso}\lesssim75\:$kpc scenarios.

\item The slope of the BHMF at low SMBH masses is found to be a crucial way to distinguish different seeding models. Our favoured cases with $d_{\rm iso}<75\:$kpc match the high-mass end of the BHMF and is in reasonable agreement with estimates at the low-mass end. However, it should be noted that the observational constraints are relatively uncertain in this regime and many lower-mass SMBHs may be missed in current surveys.

%but overestimate the number of small mass SMBHs (i.e., $\sim 10^{5-7} M_{\odot}$). 
%Recall that Paper II indicates a fiducial value of $d_{\rm iso} \sim$100 kpc based mainly on the prediction of the SMBHs number density at z$\sim$0 compared to local estimates. 
%This tension can be alleviated if we assume that observations are missing a consistent population of low mass SMBHs, which in turn would increase the estimated local SMBH number density and the slope the BHMF at the low end. 

\item The predicted $M_{\rm BH}-M_{\rm star}$ relations at redshift zero show some differences between the models. The main differences are a trend for the Pop III.1 models to have modestly steeper indices in the very high-mass regime, i.e., with $M_{\rm star}>10^{11}\:M_\odot$. Larger differences are present in the low-mass regime, i.e., with $M_{\rm star}<10^9\:M_\odot$, which reflects the imprint of the assumed mass scale of the seeds, i.e., $M_{\rm BH}= 10^5\:M_\odot$. We note that so far we have made very simple assumptions for this seed mass. We also note that the observational data in the regime are subject to significant uncertainties due to the difficulty of obtaining a complete census of SMBHs in this regime.

%generally aligns with local fits and we cannot identify a preferred seeding scheme with current observational constraints. Unfortunately, the mass range in which we could distinguish among different seeding models is still poorly sampled by observations. 

\item At high redshifts, comparison of the model $M_{\rm BH}-M_{\rm star}$ relations with observational constraints derived from JWST-detected AGN candidates remains open to debate as the probed ranges of BH and stellar masses are likely biased upward by the most luminous objects and not well sampled by the limited volume of the simulation considered in this study. 

\item The distribution of $\lambda_{\rm edd}$ suggests that, within the viscous accretion model for the BHs adopted in \gaea, massive seeds do not grow very efficiently in their early phases, while light seeds tend to accrete at the maximum rate allowed (i.e., 10 times Eddington). 

%However, none of the proposed scenarios is able to produce enough massive BHs which align with the recent JWST observations, assuming there are no systematic effects significantly biasing the data. 

\item All three local, $z\sim0$ metrics of occupation fraction as a function of the galaxy stellar mass, galaxy stellar mass function (GSMF), and black hole mass function (BHMF) suggest a constraint of $d_{\rm iso}<75\:$kpc. Such a value places a constraint on physical models for the isolation distance, e.g., due to photoionization from the Pop III.1 source.

\end{itemize}

Expanding on this final point, a reference scale for radiative feedback from the Pop III.1 sources themselves is the radius of the Str\"omgren sphere of a supermassive, $\sim 10^5\:M_\odot$ protostar, which may have a final phase of evolution that involves being on or close to the zero age main sequence (ZAMS) for several Myr. Such a star is expected to have a H-ionizing photon luminosity of $S \sim 10^{53}$ H-ionizing photons per second and to heat its HII region to temperatures of $T\sim 30,000\:$K. Then, the radius of the HII region adopting a mean intergalactic medium density is
\begin{equation}
    r_{\rm HII} = 59.6 S_{53}^{1/3} T_{3e4}^{0.27} \left(\frac{n_{\rm H}}{n_{\rm H,z=30}}\right)^{-2/3}\:{\rm kpc},
    \label{eq:rhii}
\end{equation}
where $S_{\rm 53}\equiv S/(10^{53}\:{\rm s}^{-1})$, $T_{3e4}\equiv T/(3\times 10^4\:{\rm K})$, and $n_{\rm H,z=30}$ is the mean number density of H nuclei in the IGM at $z=30$. We note that this mean density scales as $(1+z)^3$, so by $z=20$, the mean density drops by a factor of 0.310, which would increase $r_{\rm HII}$ by a factor of 2.18. We also note that the actual size of the HII region may be limited by R-type expansion, with the timescale to establish ionization equilibrium being longer than 10~Myr. This would tend to make the size of the HII region somewhat smaller than the estimate given in Eq.~\eqref{eq:rhii}. In spite of these uncertainties, we consider the close correspondence of this ionization feedback scale with the constraint on $d_{\rm iso}<75\:$kpc derived from our semi-analytic modeling of galaxy evolution and SMBH growth to indicate that this feedback process may well play an important role in setting the conditions for Pop III.1 supermassive star and SMBH formation, with the regions affected by the HII regions forming lower-mass Pop III.2 stars \citep{2006MNRAS.366..247J,2006MNRAS.373..128G} (see \S\ref{sec:intro}).

The next major step that is needed to enable further comparison of model results with observational data is to make predictions for the luminosities of the galaxies and AGN in these simulations. This is the focus of a follow-up paper to this current work (Cammelli et al., in prep.). Another promising avenue is to utilize the developed SMBH growth models to make predictions for the gravitational wave emission from these different seeding scenarios (Singh et al., in prep.).

\section*{Acknowledgments}

We thank the referee for the useful comments that helped improving the quality of this work. We also thank Jacopo Salvalaggio and Mahsa Sanati for the helpful discussions. We acknowledge the support of the computing centre of INAF-Osservatorio Astronomico di Trieste, under the coordination of the CHIPP project \citep{Bertocco19, Taffoni20}. JCT acknowledges support from ERC Advanced Grant MSTAR.

%%%%%%%%%%%%%%%%%%%%%%%%%%%%%%%%%%%%%%%%%%%%%%%%%%
\section*{Data Availability}
The data underlying this article will be shared on reasonable request to the corresponding author.


\bibliographystyle{mnras}
\bibliography{my_bib} % if your bibtex file is called example.bib

%\input{output.bbl}

%\begin{thebibliography}{99}
%\end{thebibliography}

%%%%%%%%%%%%%%%%%%%%%%%%%%%%%%%%%%%%%%%%%%%%%%%%%%

%%%%%%%%%%%%%%%%% APPENDICES %%%%%%%%%%%%%%%%%%%%%

\appendix
\section{Galaxy survival time}\label{appendix:time}

\begin{figure}
    \centering
    \includegraphics[width=\columnwidth]{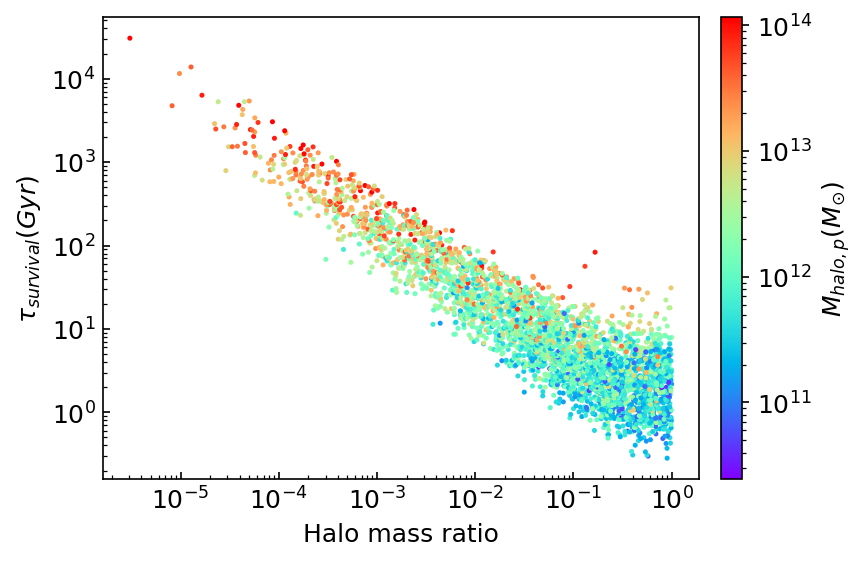}
    \caption{Estimated total survival time in Gyr for seeded halos for the $d_{\rm iso}=50\:$kpc case as a function of the halo mass ratio and color coded with the mass of the most massive (primary) halo. }
    \label{fig:merg_time}
\end{figure}
As discussed in \S\ref{sec:calibration}, we adopt the sum of two different time scales when dealing with galaxy mergers. First, we estimate the time the subhalo will survive within the main halo group, which has been evaluated via comparison against N-body simulations \citet{Berner22}. Secondly we keep track of the orphan galaxies by adding a second time which mimics the phase during which a galaxy would orbit around the central one and eventually merge with it. In Fig.~\ref{fig:merg_time} we present the estimated total survival time of satellite galaxies as a function of the halo mass ratio calculated at the halo merging time taken from \pinocchio using the values reported in Eq.~\eqref{eq:params}. In particular, we report data from the seeded halos for the $d_{\rm iso}=50\:$kpc case. This distribution of points has been calibrated to reproduce the Millennium one as obtained from the standard implementation of \gaea. The impact of the total survival time proposed in Eq.~\eqref{eq:t_sat}) and \eqref{eq:t_orph} introduces a dual dependence. Mergers with comparable halo masses tend to last on average few Gyr and typically this happens when the two merging halos belong to the low mass end of the halo mass function. As one increases the mass of the primary halo, it will eventually merge with even smaller halos. For these cases the total survival time will overshoot the age of the Universe and the two structures will never merge. This is in agreement with findings in simulations \citep{Somerville15}. The more comparable are the encounters, the more efficient we expect the interaction to be (i.e., via dynamical friction, ram pressure stripping, tidal disruption forces, etc.), resulting in a shorter time scale. Conversely, on average small galaxies joining a massive central galaxy group are thought to undergo very minor physical effects, ending up orbiting around the central galaxy undisturbed for tens or thousands of Hubble times. 
    
\section{Calibrated Galaxy Stellar Mass Function}\label{appendix:gsmf}

\begin{figure*}[B]
    \centering
    \includegraphics[width=\textwidth]{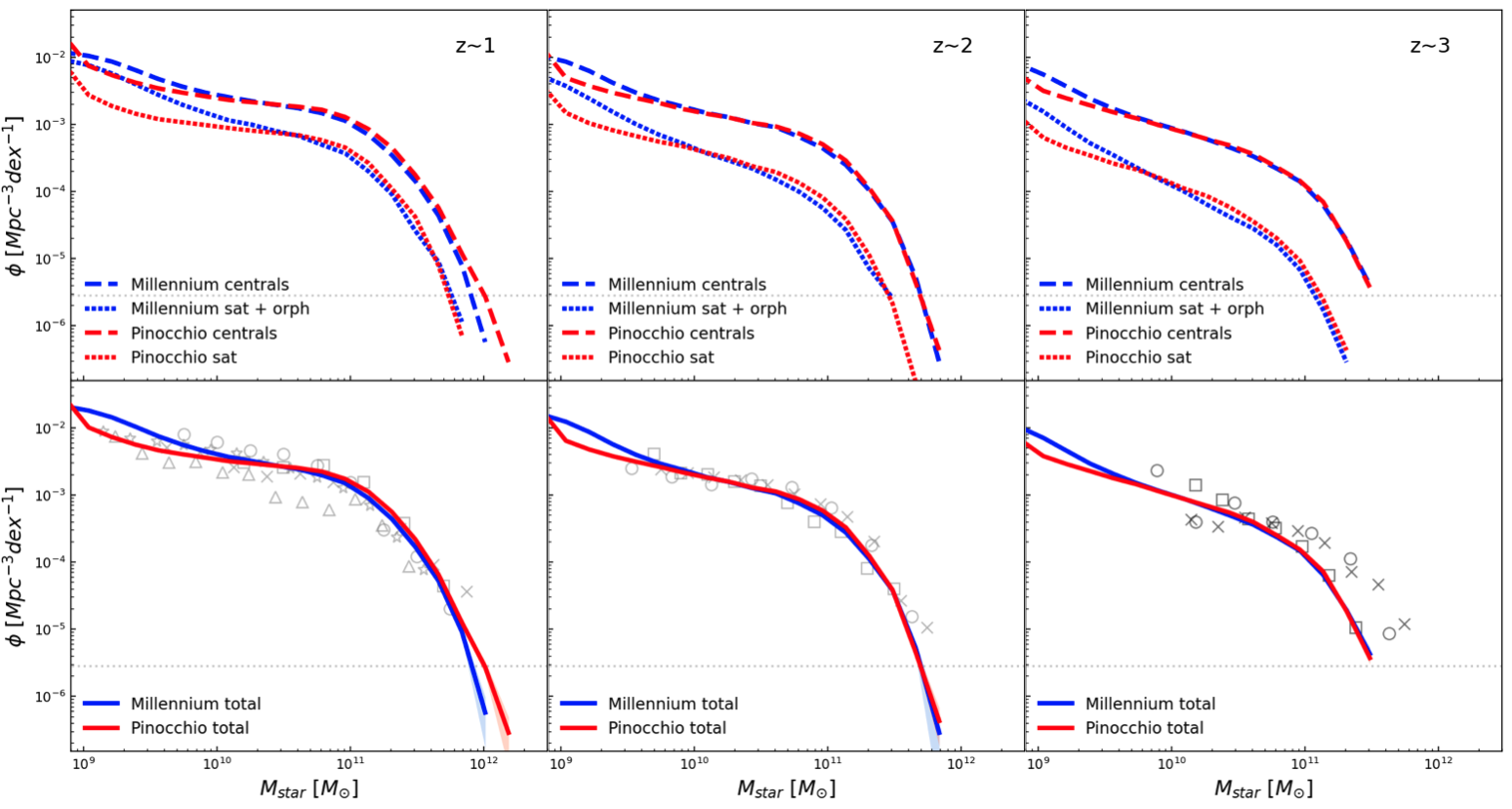}
    \caption{GSMF predictions based on the Millennium simulation box (blue lines) compared to the GSMF extracted from our Millennium-like \pinocchio box run used for the calibration of the merging times (see \S\ref{sec:calibration}). \textit{Upper panels}: satellite and central galaxies contributions to the GSMF are reported separately in dotted and dashed lines, respectively. \textit{Lower panels}: total GSMFs. Gray symbols show observational estimates as reported in Fig.~\ref{fig:smf_z0_mill}.}
    \label{fig:smf_mill}
\end{figure*}

The galaxy survival time introduced in \S\ref{sec:trees} and showed in Appendix~\ref{appendix:time} as a function of the halo mass ratio plays a crucial role in shaping the GSMF. Once again we note that our calibration process has been performed aiming at reproducing the Millennium-based result at redshift 0, and not against observational data. In this appendix we show the results from the calibrated GSMFs at redshifts $z\sim 1$ to 3. In particular we emphasize that our \pinocchio-based GSMF well reproduces the Millennium-based predictions up to $z\sim3$ and above. The agreement between the 2 different trends is depicted in Fig.~\ref{fig:smf_mill}, where we indicate Millennium- and \pinocchio-based GSMFs in blue and red, respectively. This supports the goodness of our calibration of the galaxy survival time in order to obtain reliable model galaxy catalogs.

%%%%%%%%%%%%%%%%%%%%%%%%%%%%%%%%%%%%%%%%%%%%%%%%%%

% Don't change these lines
% \bsp	% typesetting comment
\label{lastpage}
\end{document}